\documentstyle[11pt,epsfig]{article}

\def\Journal#1#2#3#4{{#1} {\bf #2} (#4) #3}

\def\NPB{{\em Nucl. Phys.} B}
\def\PLB{{\em Phys. Lett.}  B}
\def\PRL{\em Phys. Rev. Lett.}
\def\PRD{{\em Phys. Rev.} D}

\def\be{\begin{equation}}
\def\ee{\end{equation}}
\def\bea{\begin{eqnarray}}
\def\eea{\end{eqnarray}}

\begin{document}

\begin{center}
 {\Large\bf Odd-flavor Hybrid Monte Carlo Algorithm for Lattice QCD 
        }
\end{center}
\vskip 0.5 cm

\begin{center}
{\large 
  Tetsuya~Takaishi$^{\scriptscriptstyle a}$ and
  Philippe~de~Forcrand$^{\scriptscriptstyle b,c}$
}
\end{center}
\vskip 2.3ex
\begin{center}
\it
$^{\scriptstyle a}$ Hiroshima University of Economics, Hiroshima, 731-0192, JAPAN \\

$^{\scriptstyle b}$ Inst. f\"ur Theoretische
                          Physik, ETH H\"onggerberg, CH-8093 Z\"urich, Switzerland \\ 
$^{\scriptstyle c}$ CERN, Theory Division, CH-1211 Gen\`eve 23, Switzerland \\

\end{center}

\begin{abstract}

We discuss hybrid Monte Carlo algorithms 
for odd-flavor lattice QCD simulations. 
The algorithms include a polynomial approximation
which enables us to simulate odd-flavor QCD in the framework 
of the hybrid Monte Carlo  algorithm.    
In order to make the algorithms exact,
the correction factor to the polynomial approximation
is also included in an economical, stochastic way. 
We test the algorithms for $n_f=1$, 1+1 and 2+1 flavors and 
compare results with other algorithms.

\end{abstract}

\section{Introduction}

In lattice QCD simulations the standard algorithm to incorporate effects of
dynamical fermions is the Hybrid Monte Carlo (HMC) algorithm \cite{HMC}
which is conventionally used to simulate two-flavor QCD. 
Due to the algorithmic limitation of the HMC,
those simulations are limited to even numbers of degenerate flavors.
In order to include dynamical effects correctly, simulations of
QCD with three light flavors (u,d,s quarks) are desirable.
Simulations with an odd number of flavors can be performed using
the R-algorithm \cite{Ralg}. 
Also for two flavors of staggered fermions \cite{KS}, the R-algorithm is chosen
because the HMC is not applicable.
The R-algorithm, however, is not exact:
it causes systematic errors of order $(\Delta\tau)^2$, where $\Delta\tau$
is the step-size of the Molecular Dynamics evolution. A careful
extrapolation to zero step-size is therefore needed to obtain exact
results. Nevertheless, it is common practice to forego this extrapolation
and to perform simulations with a single step-size chosen small enough
that the expected systematic errors are smaller than the statistical ones.
In this paper, we wish to emphasize that there is an alternative to the R-algorithm,
which gives for our problem arbitrarily accurate results (for infinite
computer time) without any extrapolation to $\Delta \tau = 0$\cite{other}. 
Furthermore, we can make our algorithm exact with an additional, stochastic Metropolis test
which we describe and implement.

L\"uscher proposed a local algorithm, the so-called "Multiboson algorithm"
\cite{Luscher},
in which the inverse of the fermion matrix
is approximated by a suitable Chebyshev polynomial.
Originally he proposed it for two-flavor QCD. 
Bori\c ci and de Forcrand \cite{BPh} noticed that 
the determinant of a fermion matrix can be written in a manifestly positive way 
using a polynomial approximation, 
so that one can simulate odd flavors QCD with the multiboson method.
Indeed, using this method, one-flavor QCD was simulated successfully
\cite{NF1}.
The same polynomial approximation can be applied for the HMC \cite{FFMC}.
The first example of a polynomial approximation within HMC simulations of 
two-flavor QCD was made by the authors of \cite{FFMC}, and  later by \cite{Jansen}.
Application to odd flavors QCD is straightforward when one uses 
Bori\c ci and de Forcrand's idea.
Actually, in the development stage of Ref.\cite{NF1},
one-flavor QCD was also simulated by HMC and
it was confirmed that the two algorithmically different methods ---
multiboson and HMC --- give the same distribution of
plaquette values \cite{nf1HMC}.

In preliminary work\cite{NF3}, we have simulated $n_f=3$ QCD by the HMC
with a polynomial approximation,
and shown that the results are consistent with those from the R-algorithm.
However, the algorithm developed in \cite{NF3} is not exact yet.
A polynomial approximation of moderate degree $n$ may not be sufficiently accurate,
especially for small quark masses.
For such a case, it is important to control the errors coming from the polynomial approximation. 
In this study, in order to make our algorithm exact, 
we include the correction factor to the polynomial approximation 
into the algorithm and compare our results with those from other algorithms 
( R-algorithm with one flavor, and two-flavor HMC ).

A problem specific to odd flavors is the so-called "sign problem":       
the fermionic determinant may change sign configuration by configuration
and one has to include the sign into the statistical average (when quarks
come in equal-mass pairs, the determinant is squared and its sign does not
matter).
The calculation of the sign of the determinant on each configuration is 
feasible in principle, but is very costly. It is common practice to assume
that the QCD determinant is always positive. This is the case in the continuum
limit, and there are indications that it is also the case for the lattice
spacings normally considered and the quark masses currently 
reachable~\cite{Farchioni}.
Anyway, our goal here is not to study the sign problem, 
but rather to present an alternative method to the R-algorithm,
which samples the absolute value of the determinant like the R-algorithm,
but requires no extrapolation.

Our paper is organized as follows. In Sec. 2 we describe the standard HMC algorithm
which is used for even-flavor QCD simulations.
In Sec. 3 we give our algorithm to simulate odd-flavor QCD, including the additional,
stochastic Metropolis step which makes the algorithm exact.
In Sec. 4-6 we show results from our algorithm and compare them with other algorithms.
We give our conclusions in Sec. 7.

\section{Standard Hybrid Monte Carlo Algorithm}
The lattice QCD partition function with $n_f$ flavors is given by
\be
Z=\int [dU] (\Pi_{i}^{n_f} \det D_i) \exp(-S_{g}[U]),
\ee
where $D_i$ is the Wilson Dirac fermion matrix, $D_i=1-\kappa_i M[U]$ with a hopping parameter $\kappa_i$
and $M[U]$ is the lattice Wilson hopping operator.  
$S_{g}[U]$ stands for the gauge action. 

The Hybrid Monte Carlo (HMC) algorithm\cite{HMC} was developed for
a system containing  multiples of two degenerate quark flavors.
For instance, for $n_f=2$ flavors ( two degenerate quarks ), one has the partition function
\be 
Z=\int [dU] \det D^2 \exp(-S_{g}[U]).
\ee
Using a pseudofermion field $\phi$, an integral representation of the determinant factor $\det D^2$  can be given by
\be
\det D^2 \sim \int [d{\phi}^\dagger] [d{\phi}]
\exp(- {\phi}^\dagger D^{\dagger -1}D^{-1} \phi),
\label{eq:intnf2}
\ee
where the relation, $\det D^\dagger =\det D$, is used.  
The term ${\phi}^\dagger {D}^{\dagger -1}{D}^{-1}\phi$ is written in a manifestly positive form
which is essential to formulate the HMC algorithm.
Introducing momenta $P_i$ conjugate to the link variables $U_i$, the partition
function is rewritten as
\be
Z=\int [dU] [dP] [d{\phi}^\dagger] [d{\phi}] \exp(-H),
\ee
where the Hamiltonian $H$ is defined by
\be
H=\frac{1}{2}P^2 +S_{g}[U]   +  {\phi}^\dagger {D}^{\dagger -1}{D}^{-1} {\phi},
\label{eq:standard-H}
\ee
where $P^2 \equiv \sum_i P^2_i$.

The HMC algorithm consists of two steps: molecular dynamics (MD) evolution and Metropolis accept/reject step.
In the MD evolution, one solves Hamilton's equations of motion:
\bea
\frac{d U_i}{d t} & = & \frac{\partial H}{\partial P_i} \\ 
\frac{d P_i}{d t} & = & -\frac{\partial H}{\partial U_i}.
\eea
In general these equations are not  solvable analytically.
Usually one solves the equations approximately by the 2nd order leapfrog integrator 
which is time-reversible and area-preserving, thanks to which detailed balance is satisfied.
After integrating the equations, one obtains a candidate configuration $(U^\prime,P^\prime)$ 
from the starting configuration $(U,P)$.
Next, one performs a Metropolis test                                           
with  acceptance probability $p=\min(1,\exp(-\delta H))$
where $\delta H=H(U^\prime,P^\prime)-H(U,P)$.
In case of rejection, one keeps the old $U$.
In this study we call the HMC algorithm with Hamiltonian eq.(\ref{eq:standard-H}) the standard HMC algorithm.

\section{Algorithm for Odd Number of Flavors}
\subsection{$n_f=1$ algorithm}
\label{sec:nf1}

The partition function of  $n_f=1$ QCD is given by
\be
Z=\int [dU] \det D \exp(-S_{g}[U]).
\ee
In order to simulate $n_f=1$ QCD, 
one has to treat a single $\det D$ which can not be incorporated into the standard HMC.
Bori\c ci and de Forcrand \cite{BPh} noticed that a single $\det D$, when positive, can be
written in a manifestly positive way when  L\"uscher's polynomial approximation\cite{Luscher} is used.

One can approximate the inverse of $D$ by a polynomial,
\be
1/D\approx P_m(D) \equiv \prod_{k=1}^{m}(D-z_k),
\label{eq:Pn}
\ee
where $z_k$ are the roots of the polynomial $P_m(D)$ (a common choice is
$z_k=1-\exp(i~2\pi k/(m+1))$).
For a polynomial of even degree $m=2n$, the roots come in complex conjugate pairs
$(z_{2k-1},z_{2n+2-2k}), k=1,..,n)$.
Thus, eq.(\ref{eq:Pn}) is rewritten as
\be
P_{2n}(D) = \prod_{k=1}^{n}(D-z_{2k-1})(D-\bar{z}_{2k-1})
\ee
where $z_k=1-\exp(i~2\pi k/(2n+1))$.
Using the $\gamma_5$ hermiticity of the fermion matrix,
one finds that $\det (D-\bar{z}_k)=\det(D-z_k)^\dagger$.
Introducing  $T_n(D)\equiv\prod_{k=1}^n(D-z_{2k-1})$,
the determinant of $D$ is  written as
\be
\det (D) = C_n\det ( T_n^\dagger(D)T_n(D))^{-1}, 
\ee
with the correction factor $C_n$ given by
\be
C_n \equiv \det(DT_n^\dagger(D)T_n(D)) 
\label{Cn}
\ee
which goes to one in the limit $n \rightarrow \infty$.
An integral representation of  $\det ( T_n^\dagger(D)T_n(D))^{-1}$ is given by 
\be
\det (T_n^\dagger(D)T_n(D))^{-1} \sim \int [d\phi^\dagger] [d\phi] 
\exp(-\phi^\dagger T_n^\dagger(D)T_n(D)\phi).
\label{eq:intnf1}
\ee
Note that the term $\phi^\dagger T_n^\dagger (D)T_n(D)\phi$  
is Hermitian positive, which can not be 
realized for one flavor in the standard HMC formulation.

Now, the $n_f=1$ Hamiltonian for our HMC algorithm can be defined as 
\be
H^{n_f=1} =\frac{1}{2}P^2 +S_{g}[U]  + \phi^\dagger  T_n^\dagger(D)T_n(D)\phi.
\label{eq:nf1-H}
\ee
Making use of this Hamiltonian one can perform HMC simulations for $n_f=1$ QCD.

The Hamiltonian defined by eq.(\ref{eq:nf1-H}) is an approximation to the exact one, 
which generates configurations with the measure $\sim \det( T_n^\dagger(D)T_n(D))^{-1}$.
The quality of this approximation is measured by the deviation of $C_n$ (eq.(\ref{Cn})) from one.
To make the algorithm exact, one has to include the correction factor $C_n$ into the algorithm.
There are several possibilities to incorporate $C_n$ in the measure\cite{Forcrand99}.
One possibility is to make a global Metropolis test with probability
\be
p=\min\left[1,\frac{C_n^\prime}{C_n}\right],
\label{met}
\ee
where $C_n^\prime\equiv C_n[U^\prime]$ and $C_n \equiv C_n[U]$ correspond to a new candidate configuration $U^\prime$ 
and a starting configuration $U$ respectively.
Since the correction factor is the determinant of a large matrix,
its direct calculation is not feasible\footnote{In \cite{Forcrand95}
the correction factor was computed exactly by the Lanczos method. This approach is limited to small lattices.}.
An economical way is to form an unbiased estimator of the ratio ${C_n^\prime}/{C_n}$ 
and to use a noisy Metropolis test\cite{Kennedy85,Bhanot85}.
The same correction factor appears in the $n_f=1$ multiboson algorithm\cite{NF1},
where the ratio was estimated by rewriting the correction factor 
using another high-quality polynomial $T_r(D)$ ($r>>n$)\cite{NF1,Montvay} as
\be
C_n=\lim_{r\rightarrow \infty} \frac{\det{T_n^\dagger(D)T_n(D)}}{\det{T_r^\dagger(D)T_r(D)}}.
\ee
Using this form of the correction factor, the ratio ${C_n^\prime}/{C_n}$  is written as 
\bea
\frac{C_n^\prime}{C_n}& \equiv & \det\left[\frac{X^{\prime \dagger} X^\prime}{X^\dagger X}\right] \nonumber \\
 & = &  \frac{\int d\eta^\dagger d\eta \exp(-|X^{\prime -1}X\eta|^2)}
             {\int d\eta^\dagger d\eta \exp(-|\eta|^2)}  \nonumber \\
 & = & <\exp(-|X^{\prime -1}X\eta|^2+|\eta|^2)>_{\eta} 
\label{eq:ratio1}
\eea
where $X=T_n(D)T_r(D)^{-1}$. 
Thus the ratio can be estimated by calculating 
$\exp(-|X^{\prime -1}X\eta|^2+|\eta|^2)$  with only one Gaussian random vector $\eta$. 
However convergence is slow, and there may be a technical difficulty using a high-quality polynomial:
the above estimation includes
a number of " matrix($D$) $\times$ vector +  vector " type calculations, which results in divergence for 
high-degree polynomials due to the roundoff errors of computers\cite{NF3}.

We solve this problem very economically here, by proposing to estimate 
the ratio ${C_n^\prime}/{C_n}$ from unbiased estimators of $({C_n^\prime}/{C_n})^2$. 

$({C_n^\prime}/{C_n})^2$ is easily estimated  by
\bea
\left(\frac{C_n^\prime}{C_n}\right)^2 & = & 
\det \frac{D'^2 P_{2n}(D')^\dagger P_{2n}(D')}{D^2 P_{2n}(D)^\dagger P_{2n}(D)} = 
  \det \left(\frac{W^{\prime}}{W}\right)^2 \nonumber \\  
      & = & \frac{\int d\eta^\dagger d\eta \exp(-|W^{\prime -1}W\eta|^2)} 
                 {\int d\eta^\dagger d\eta \exp(-|\eta|^2)}                             \nonumber \\
      & = &   <\exp(-|W^{\prime -1}W\eta|^2+|\eta|^2)>_{\eta}
\label{eq:ratio2}
\eea 
where $W\equiv D P_{2n}(D)$.
Then the problem is to estimate $({C_n^\prime}/{C_n})$ 
using unbiased estimators of $({C_n^\prime}/{C_n})^2$ only.
This can be accomplished by evaluating stochastically the Taylor expansion of $\sqrt{x}$
as shown in \cite{LinLiu} for $e^x$.
Expanding about $x=1$, one writes
\be
\sqrt{1+\epsilon}=1+\frac12\epsilon +\sum_{k=2} c_k (-)^{k-1} \epsilon^k
\label{eq:stoch}
\ee
where $c_k = \frac{2k-3}{2k}c_{k-1}\in ]0,1[$. 
First, the left-hand side is assigned value 1. 
Next, with probability $\frac12$ one computes a stochastic estimator of $\epsilon_1 = x-1$ 
and adds it to the left-hand side. Then, with probability $\frac{2k-3}{2k} c_{k-1}|_{k=2}=\frac14$ 
one computes a second stochastic estimator of $\epsilon_2$ and 
adds $-\epsilon_1 \epsilon_2$  to the left-hand side, and so on. 
In our case, $x=({C_n^\prime}/{C_n})^2$.
When the procedure terminates, one obtains an unbiased estimator of $\sqrt{x}$.
There is a difficulty if the estimator becomes negative: it cannot be used as a probability
in the Metropolis test (eq.(\ref{met})).
Such positivity violations can be reduced by increasing the degree $n$,
to the point where they are never observed during the whole simulation. 

Our algorithm is summarized as follows.
\begin{enumerate}
\item  HMC: we perform Molecular Dynamics with the approximate Hamiltonian eq.(\ref{eq:nf1-H}) and  
obtain a candidate configuration $U^\prime$. 
Then we do a Metropolis test with acceptance probability $p=\min(1,\exp(-\delta H))$
\footnote{Here, as well as in the HMC algorithm, one could use the Glauber
acceptance probability $p=(1 + \exp(-\delta H))^{-1}$ instead of the Metropolis
one. This does not seem to be a judicious choice however, since even for
an exact MD evolution ($\delta H = 0$) the acceptance will only be $1/2$.}.
This removes the step-size integration error.

\item  If the candidate configuration is accepted,  we estimate the ratio ${C_n^\prime}/{C_n}$ 
stochastically by using the method explained above.  

\item  We perform another Metropolis test with acceptance probability $p=\min(1,\sqrt{x})$,
where $\sqrt{x}$ is an unbiased estimator of ${C_n^\prime}/{C_n}$.
This removes the error of the polynomial approximation to $\det D$.

\item  If accepted, we take $U^\prime$ as a new configuration. Otherwise we keep the old configuration. 

\end{enumerate}

This algorithm samples the measure $\sqrt{\det^2 D} = |\det D|$, just like the R-algorithm,
but does so in a more efficient way which avoids the extrapolation to zero step-size of the latter.
For very small quark masses in the Wilson formulation, the Dirac determinant may become negative
for some background gauge fields. This happens for a small fraction of the gauge ensemble, which
goes to zero in the continuum limit. Therefore, sampling the measure $|\det D|$ does not affect
the continuum limit of the lattice results. Nevertheless, it is possible, if desired, to correct
for negative determinants by multiplying the contribution of the corresponding configurations by
$-1$ in the Monte Carlo ensemble. This requires identifying possible negative real Dirac eigenvalues
in the configurations of that ensemble, which can be done e.g. with the Arnoldi method.

Note that the choice of approximating polynomial is in principle arbitrary: the polynomial approximation
error is removed at Step 3 above. However, a poor choice of polynomial, e.g. whose domain of approximation
does not cover the complete spectrum of the Dirac operator, will result in difficulty maintaining
positivity of the estimator of ${C_n^\prime}/{C_n}$, and in longer autocorrelation times.

\subsection{$n_f=2+1$ algorithm}
\label{sec:nf2+1}

The partition function of $n_f=2+1$ QCD is given by
\be
Z=\int [dU] \det D_k^2 \det D_l \exp(-S_{g}[U]).
\ee
This system consists of two degenerate quark flavors with a hopping parameter $\kappa_k$ and
one flavor with $\kappa_l$. 
An integral representation of the determinant factor $\det D_l$ is obtained
using the polynomial approximation as in eq.(\ref{eq:intnf1}),
whereas $\det D_k^2$ is expressed as in eq.(\ref{eq:intnf2})
\be
\det D_l \sim {\det}^{-1}( T_n^\dagger(D_l)T_n(D_l)) \sim \int [d\phi_l^\dagger] d\phi_l] 
\exp(-\phi_l^\dagger T_n^\dagger(D_l)T_n(D_l)\phi_l),
\label{eq:det2-nf2+1}
\ee
and 
\be
\det D_k^2 \sim \int [d{\phi_k}^\dagger] [d{\phi_k}]
\exp(- \phi_k^\dagger D_k^{\dagger -1}D_k^{-1} \phi_k).
\label{eq:det1-nf2+1}
\ee
Combining eq.(\ref{eq:det2-nf2+1}) and (\ref{eq:det1-nf2+1}) 
one can define the following Hamiltonian,
\be
H^{n_f=2+1} =\frac{1}{2}P^2 +S_{g}[U]  + 
\phi_k^\dagger D_k^{\dagger -1} {D_k}^{-1} \phi_k +\phi_l^\dagger T_n^\dagger(D_l)T_n(D_l)\phi_l.
\label{eq:H1-nf2+1}
\ee
Alternatively, one can express the determinant factors using only one vector $\phi$ as
\be
\det D_k^2 \det D_l \sim \int  [d\phi^\dagger] [d{\phi}] 
\exp(-\phi^\dagger  D_k^{\dagger -1} T_n^\dagger(D_l) T_n(D_l) D_k^{-1}\phi ).
\ee
Then one obtains another Hamiltonian
\be 
H^{n_f=2+1} =\frac{1}{2}P^2 +S_{g}[U]  +
\phi^\dagger  D_k^{\dagger -1} T_n^\dagger(D_l) T_n(D_l) D_k^{-1}\phi.
\ee
Both definitions of Hamiltonian can be used for HMC simulations.
In this study we take eq.(\ref{eq:H1-nf2+1}).

The correction factor to the $n_f=2+1$ Hamiltonian is given by 
\be
C_n = \det(D_lT_n^\dagger(D_l)T_n(D_l)).
\ee
This factor can be included in the same way as explained for the $n_f=1$ algorithm. 

\subsection{$n_f=1+1$ algorithm}
\label{sec:nf1+1}

The partition function of $n_f=1+1$ QCD is given by
\be
Z=\int [dU] \det D_k \det D_l \exp(-S_{g}[U]).
\ee
This system consists of two non-degenerate quark flavors 
with hopping parameters $\kappa_k$ and $\kappa_l$.
Each determinant factor can be expressed in terms of pseudofermion fields 
using polynomial approximations as 
\be
\det D_k \sim  \int [d\phi_k] [d\phi_k^\dagger] \exp(- \phi_k^\dagger T_n^\dagger(D_k)  T_n(D_k)\phi_k),
\ee
\be
\det D_l \sim  \int  [d\phi_l] [d\phi_l^\dagger] \exp( -\phi_l^\dagger  T_m^\dagger(D_l)T_m(D_l)\phi_l).
\ee
Then one defines the $n_f=1+1$ Hamiltonian as
\be
H^{n_f=1+1} =\frac{1}{2}P^2 +S_{g}[U]  
+ \phi^\dagger_k  T_n^\dagger(D_k)T_n(D_k)\phi_k + \phi^\dagger_l  T_m^\dagger(D_l)T_m(D_l)\phi_l 
\label{eq:H1-nf1+1}
\ee

Using one $\phi$ field only,
we can express the two determinant factors at once as

\be
\det D_k \det D_l \sim \int [d\phi] [d\phi^\dagger] 
\exp(- \phi^\dagger  T_n^\dagger(D_k)T_m^\dagger(D_l)T_m(D_l) T_n(D_k)\phi).
\ee
This expression results in the following Hamiltonian,
\be
H^{n_f=1+1} =\frac{1}{2}P^2 +S_{g}[U]  
 + \phi^\dagger  T_n^\dagger(D_k)T_m^\dagger(D_l)T_m(D_l) T_n(D_k)\phi.
\ee
In this study we use the definition of eq.(\ref{eq:H1-nf1+1}) for HMC simulations.
The correction factor of $n_f=1+1$ Hamiltonian is given by 
\be
C_{nm} \equiv \det\left[D_kT_n^\dagger(D_k)T_n(D_k)\cdot D_l T_m^\dagger(D_l)T_m(D_l)\right].
\ee

\section{Numerical results for $n_f=1+1$}

We have tested the above algorithms, using relatively small lattices in
order to obtain results of sufficient accuracy to expose tiny systematic
errors caused by a low-order polynomial approximation or 
a finite step-size (for the R-algorithm).

\label{sec:NRnf1+1}

\begin{figure}[ht]
\vspace{-3cm}
  \begin{center}
    \epsfig{width=10cm,file=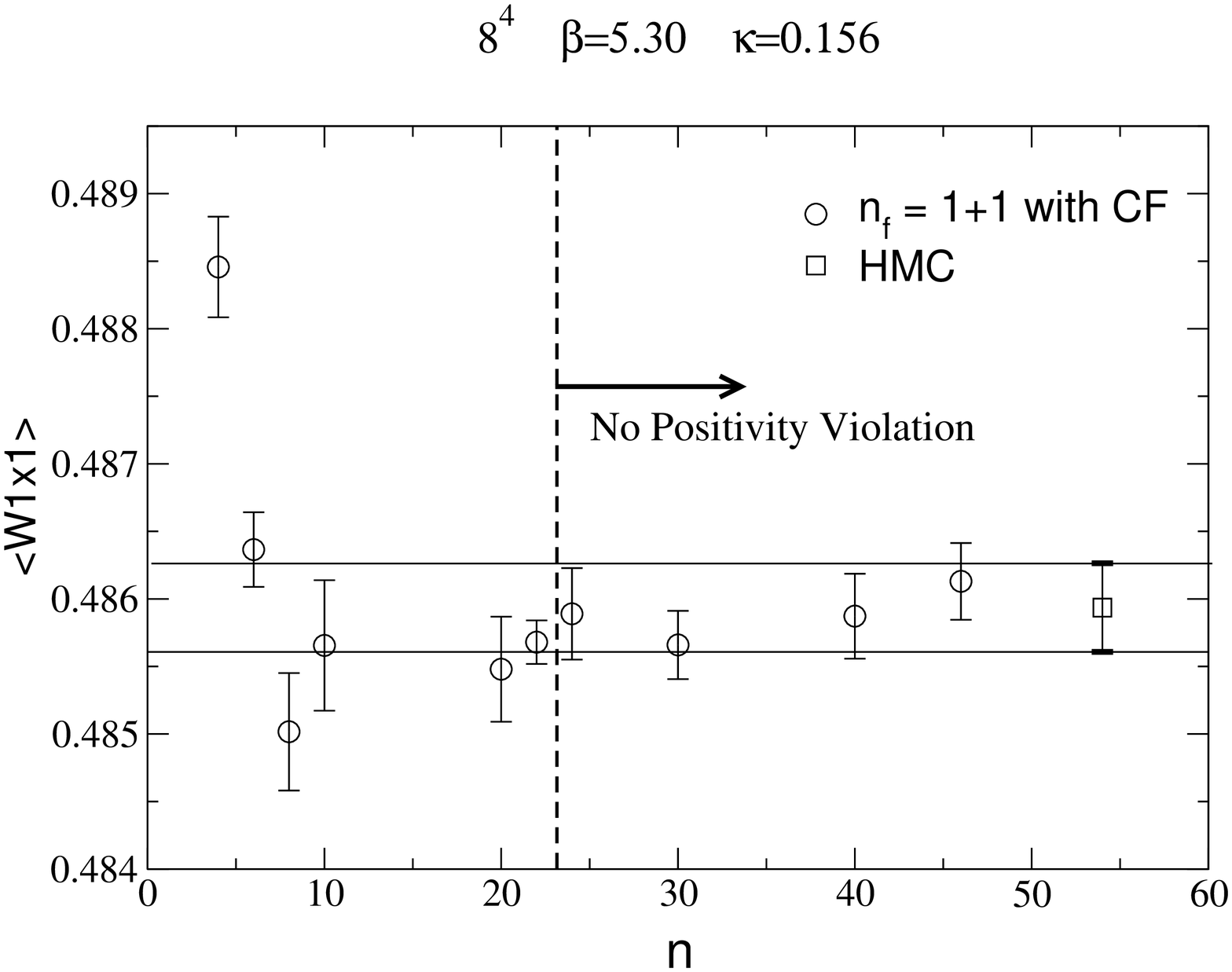}
    \epsfig{width=10cm,file=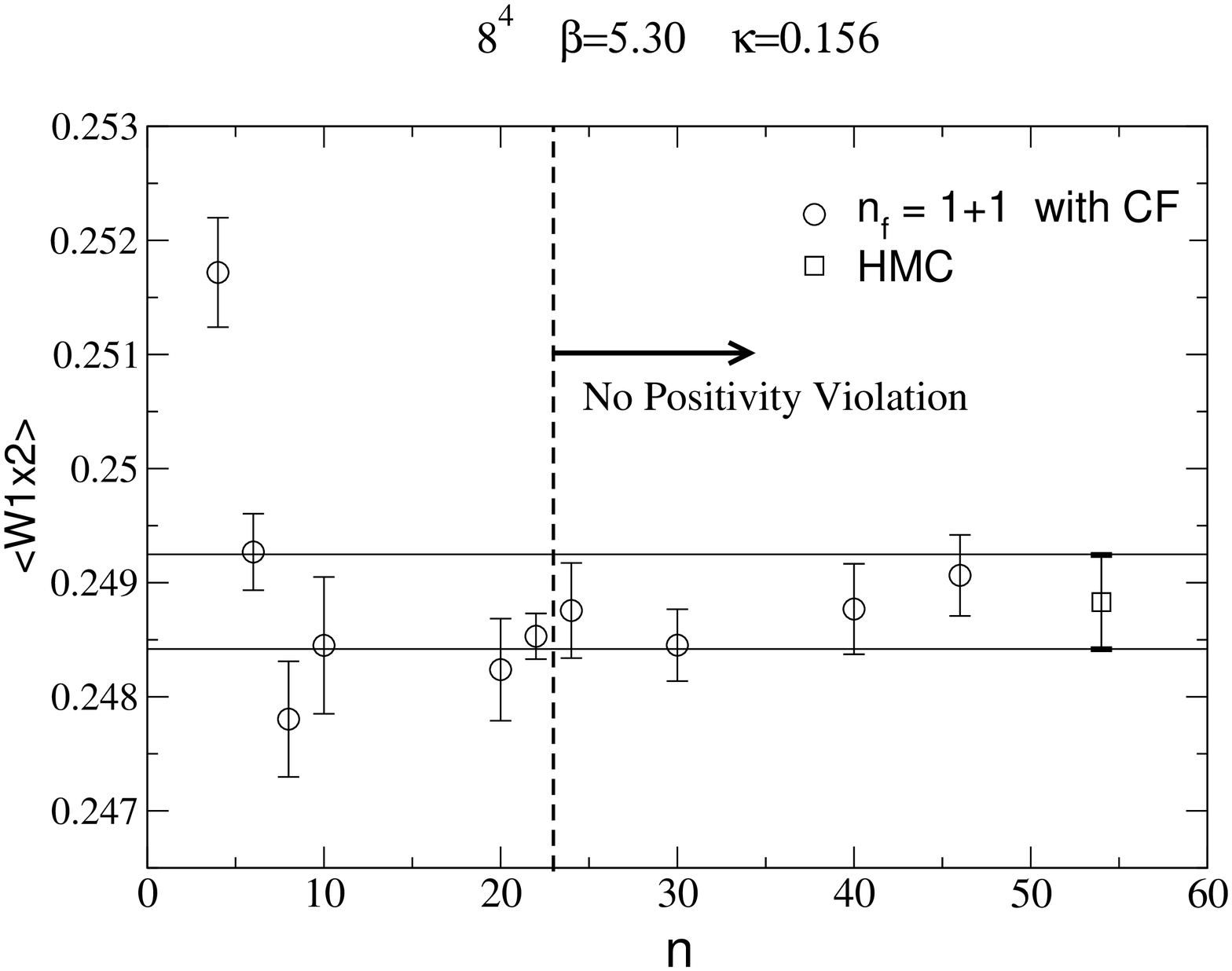}
    \epsfig{width=10cm,file=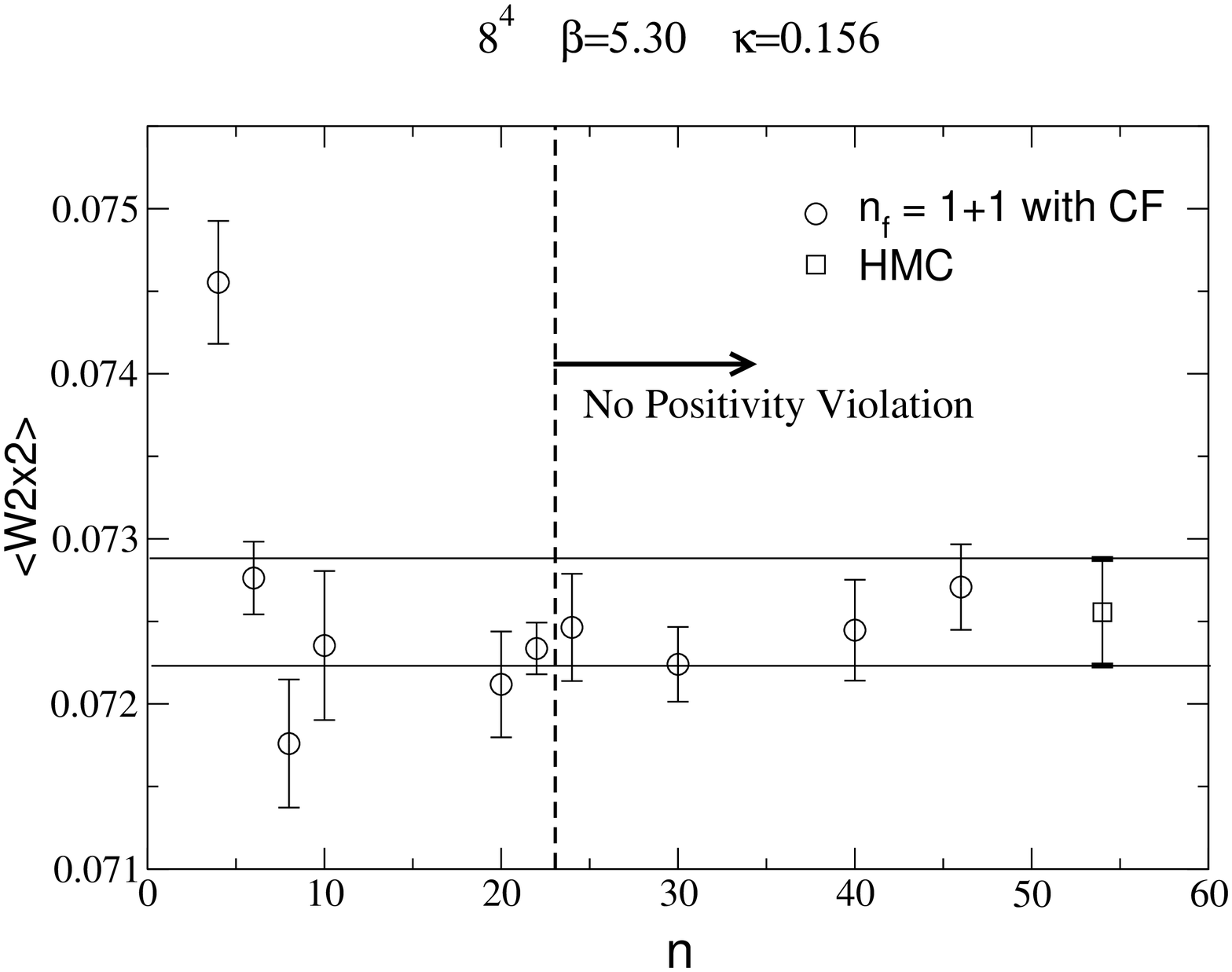}
  \end{center}
\vspace{-1cm}
\caption{$n_f=1+1$ results of Wilson loop on an $8^4$ lattice at $\beta=5.30$ and $\kappa=0.156$.}
\label{b530k156nf1+1w1x1bw}
\end{figure}

\begin{figure}[ht]
\vspace{-3cm}
  \begin{center}
    \epsfig{width=10cm,file=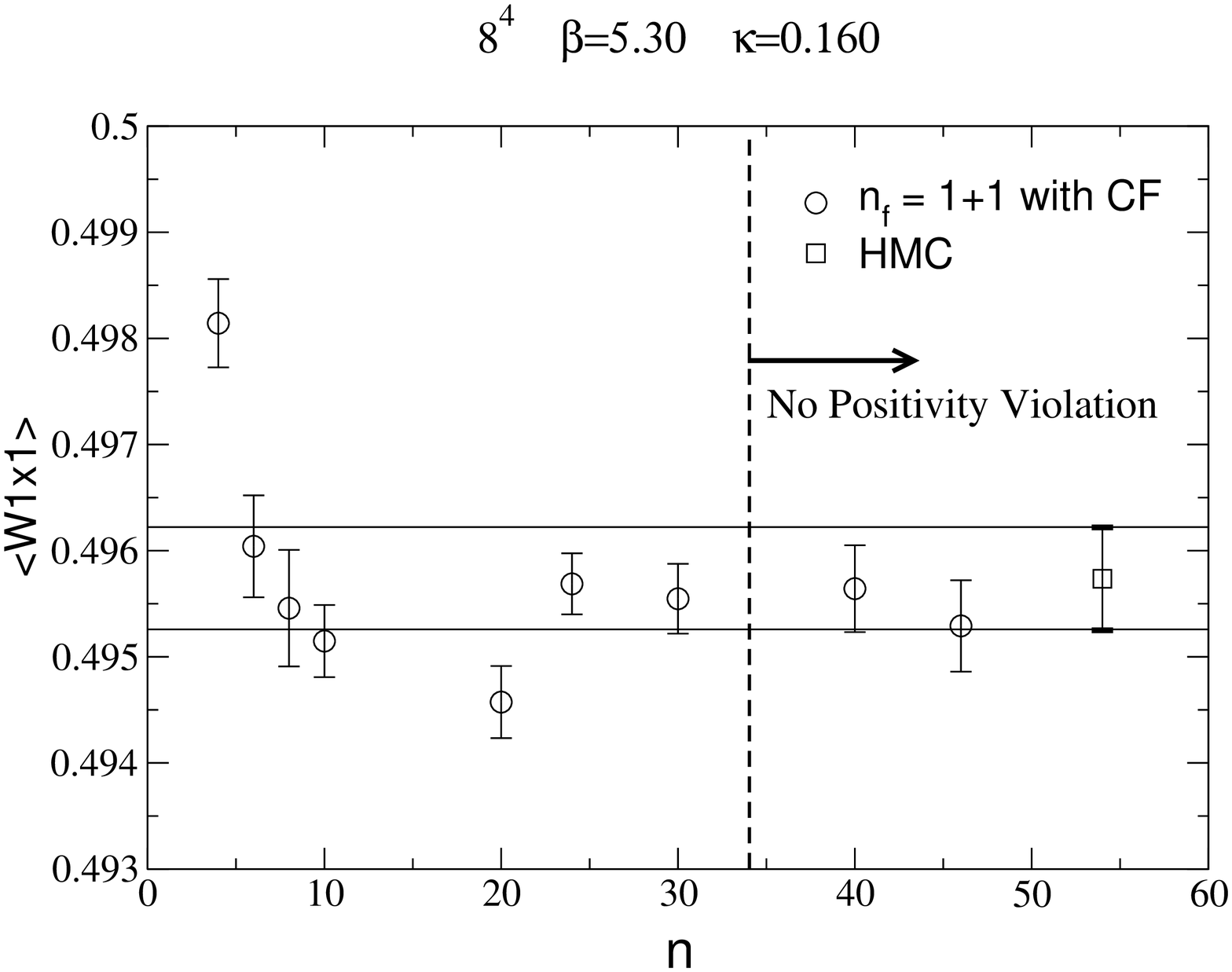}
    \epsfig{width=10cm,file=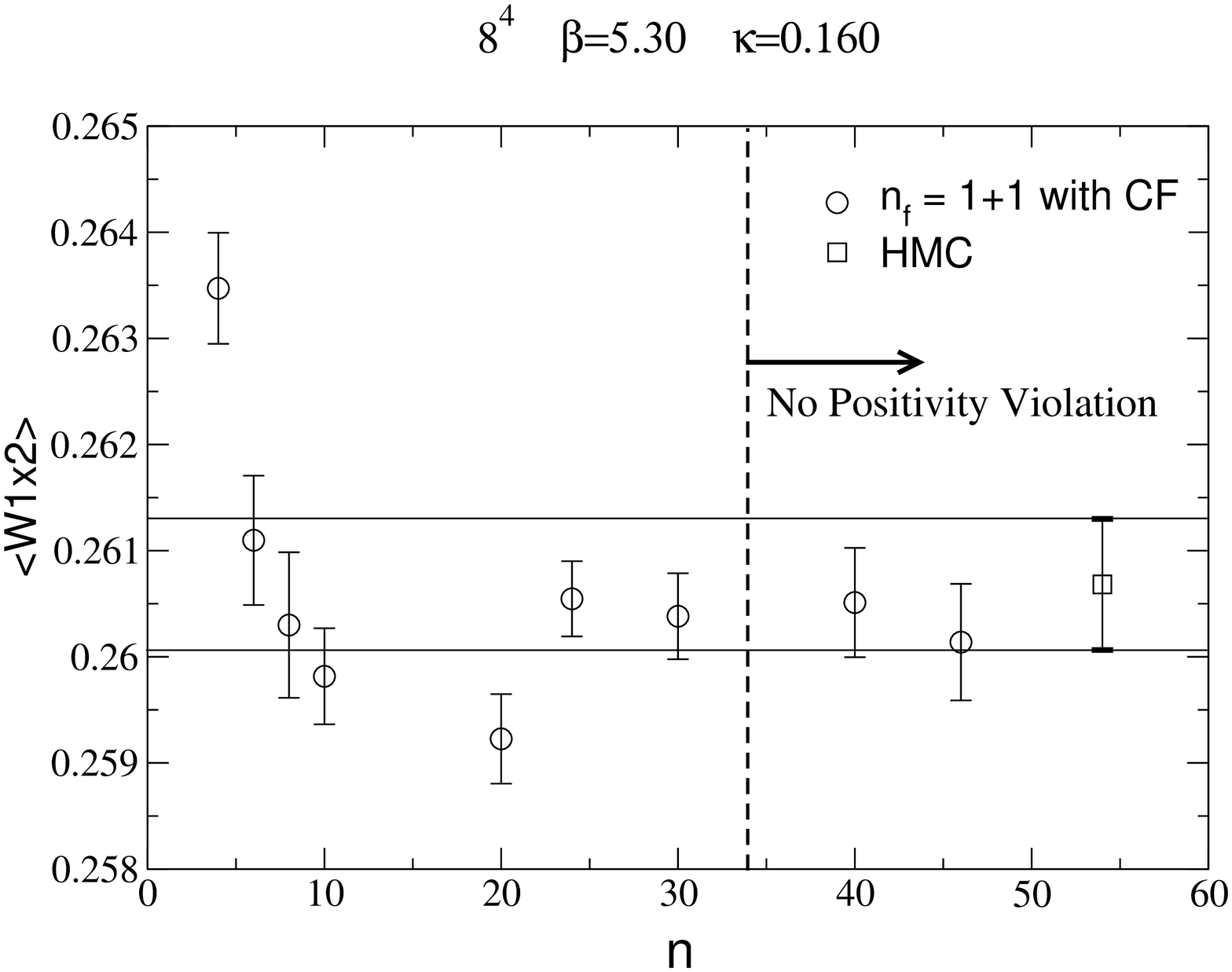}
    \epsfig{width=10cm,file=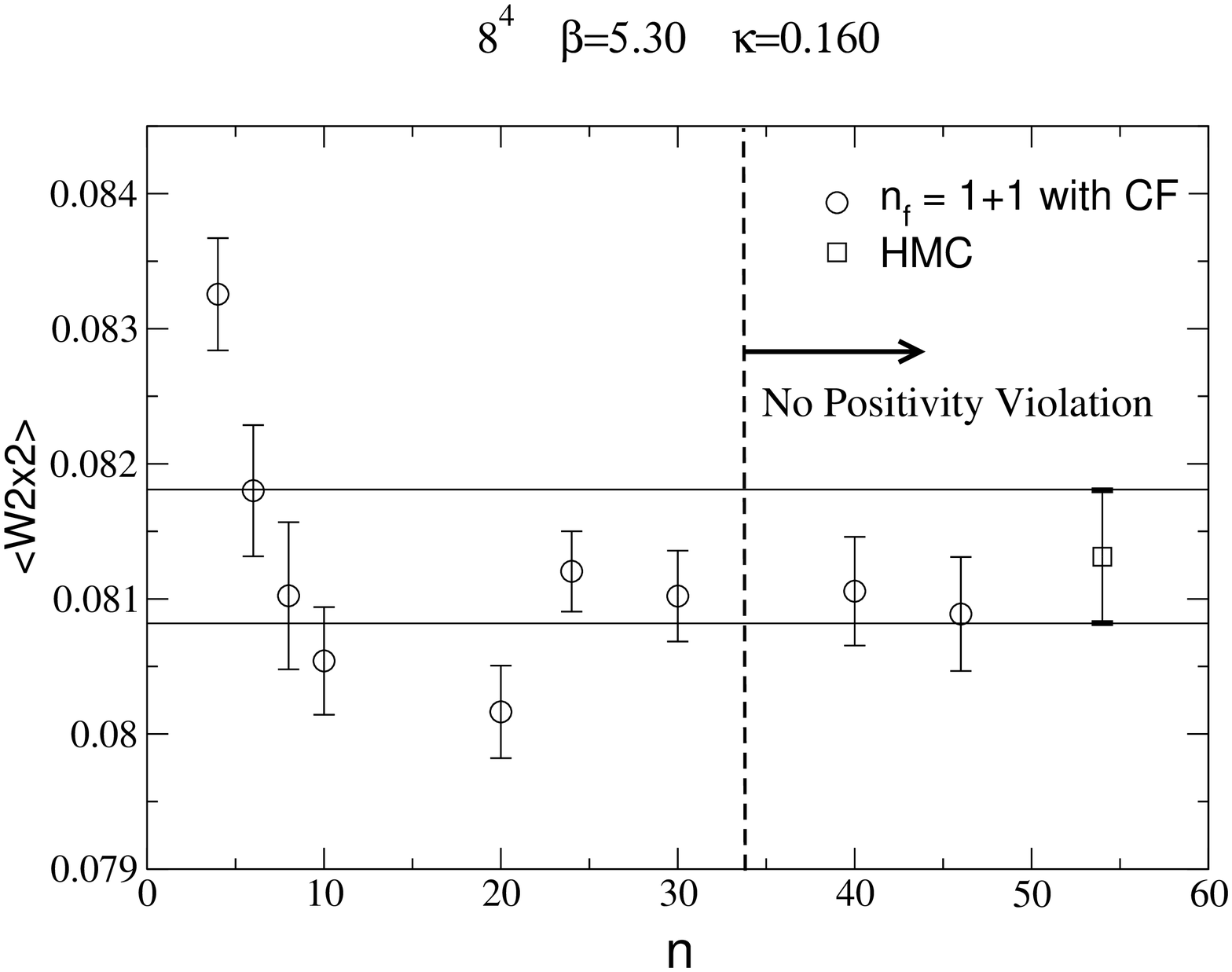}
  \end{center}
\vspace{-1cm}
\caption{Same as in fig.\ref{b530k156nf1+1w1x1bw} but for $\kappa=0.160$.}
\label{b530k160nf1+1w2x2bw}
\end{figure}

We first show results of $n_f=1+1$ QCD with {\it degenerate} quark masses.
This allows a direct comparison of our results with those of standard $n_f=2$ HMC.
The algorithm we use here is the one discussed in Sec.\ref{sec:nf1+1}, 
where we set
$\kappa\equiv \kappa_k =\kappa_l$.
We use an $8^4$ lattice at $\beta=5.30$ and $\kappa=0.156, 0.158$ and 0.160.
Boundary conditions are anti-periodic in all directions.
Since the two quarks have the same mass, we use identical polynomials
with the same degree $n$ for 
the polynomial approximation, ie. $T_n(D)=T_m(D)$ and $n=m$
and write the correction factor as $C_n\equiv C_{nm}$. 
We measure $1\times1$, $1\times2$ and $2\times2$ Wilson loops 
and vary the degree $n$ of $T_n(D)$.
The average values are taken over about 2000-8000 trajectories. 
Details are shown in Table \ref{paranf1+1}.
Average values of Wilson loops  are displayed in 
Figs.\ref{b530k156nf1+1w1x1bw}-\ref{b530k160nf1+1w2x2bw}
with those from the HMC.
Results of Wilson loops are consistent with those from the standard HMC 
except for small $n$.
These results suggest that we do not need to take a very large $n$. 
Quantitatively, however, from results of Wilson loops only it is not clear which $n$ should be chosen.  

\begin{figure}[ht]
  \begin{center}
    \epsfig{width=11cm,file=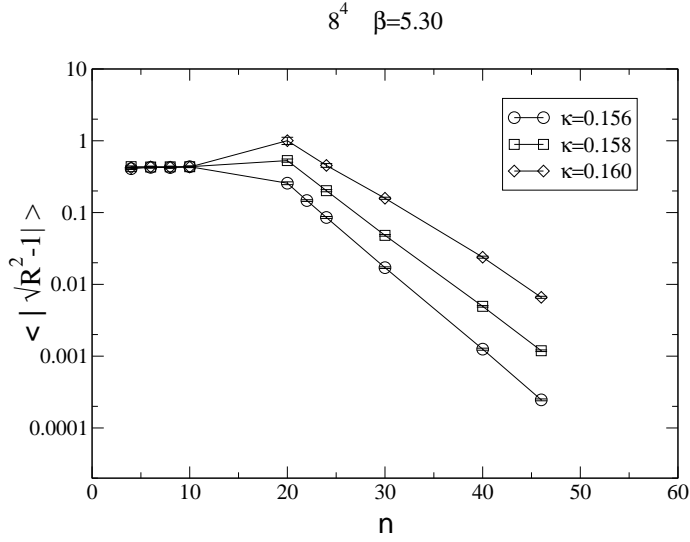}
  \end{center}
\vspace{-1cm}
\caption{$|C_n^\prime/C_n-1|$ as a function of degree $n$ 
for $n_f=1+1$ at $\kappa=0.156$, 0.158 and 0.160.
$R$ in the figure stands for $C_n^\prime/C_n$.
$C_n^\prime/C_n$ are estimated stochastically by eq.(\ref{eq:ratio2}) 
and negative values are not used in the average.}
\label{Sqb530nf1+1bw}
\end{figure}

Fig.\ref{Sqb530nf1+1bw} shows the average value of $|C_n^\prime/C_n-1|$ as a function of the degree $n$.
Negative values of $C_n^\prime/C_n$ observed for small $n$ are not included in this average.
Except for small $n$, $C_n^\prime/C_n$ converges to one exponentially as $n$ increases.
The plateau seen at small $n$ is due to the following reason.
When the polynomial approximation is inaccurate, i.e. for small $n$, 
$|W^{\prime -1}W\eta|^2$ takes a large value, and the estimated value of  $({C_n^\prime}/{C_n})^2$ via
eq.(\ref{eq:ratio2}) will always be very small. 
If we apply eq.(\ref{eq:stoch}) with $({C_n^\prime}/{C_n})^2=0$ (i.e. $\epsilon=-1$), 
we obtain $\langle {C_n^\prime}/{C_n} \rangle=0.429$ when the average excludes negative values. 
This value is consistent with our results. Anyhow we are not interested in such a small $n$ 
and should increase $n$ until no negative value of $C_n^\prime/C_n$ is observed.   

\begin{figure}[ht]
  \begin{center}
    \epsfig{width=11cm,file=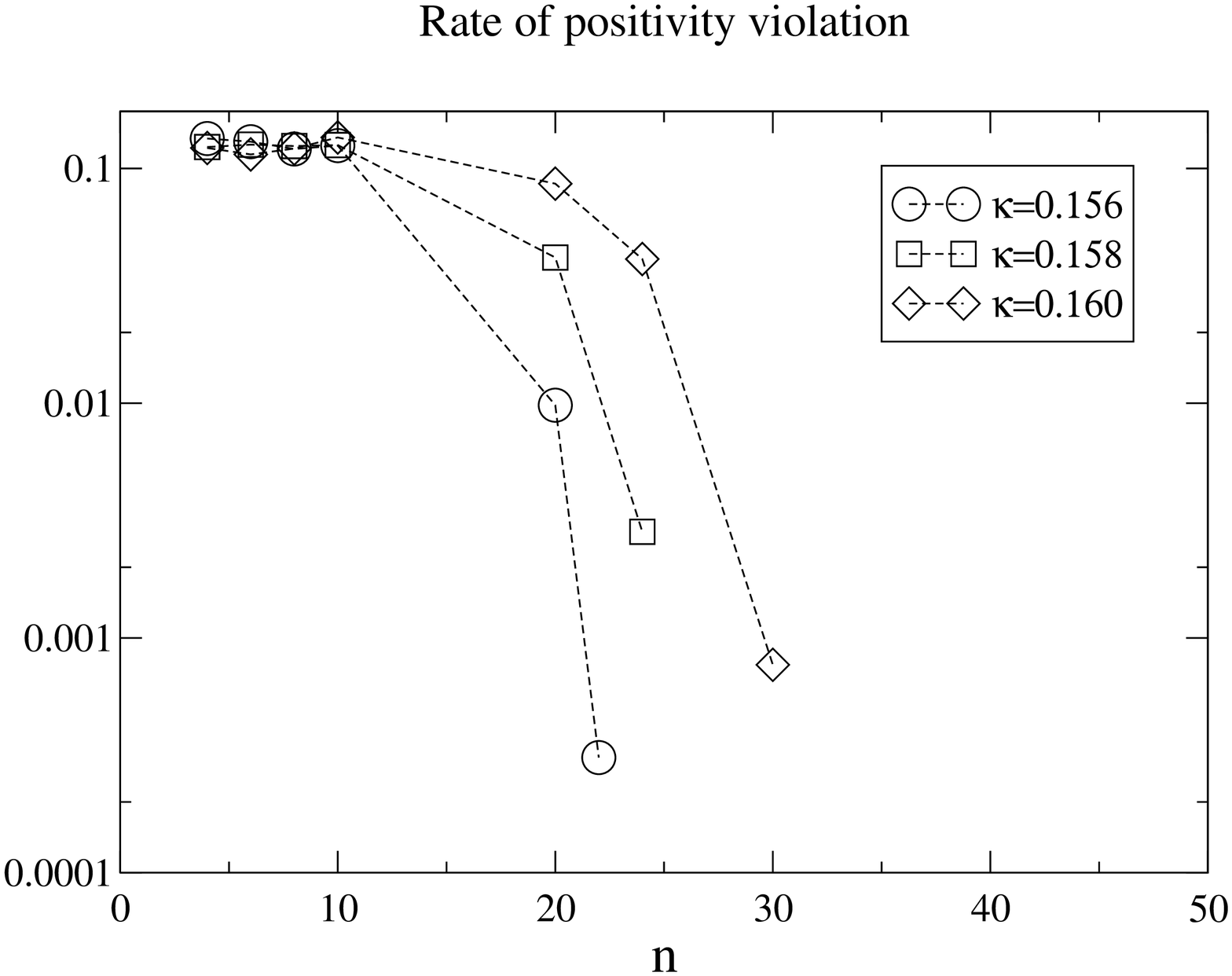}
  \end{center}
\vspace{-1cm}
\caption{Rate of positivity violation of $C_n^\prime/C_n$ for $n_f=1+1$
as a function of degree $n$.}
\label{b530L8PVbw}
\end{figure}

\begin{figure}[ht]
  \begin{center}
    \epsfig{width=11cm,file=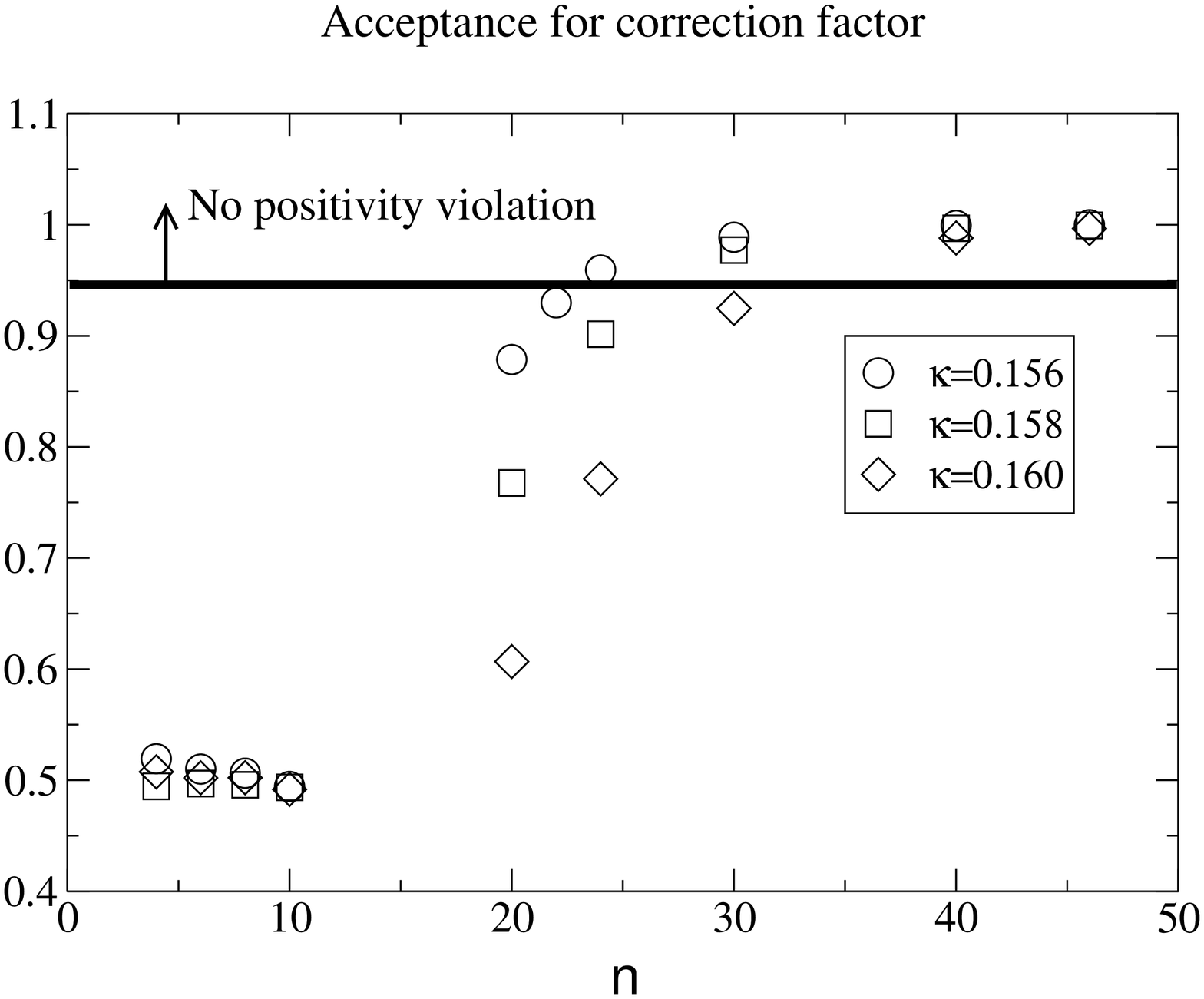}
  \end{center}
\vspace{-1cm}
\caption{Acceptance of the Metropolis test with $p=\min(1,C_n^\prime/C_n)$ for $n_f=1+1$.
}
\label{b530L8nf1+1ACbw}
\end{figure}

Fig.\ref{b530L8PVbw} shows the positivity-violation (PV) rate of $C_n^\prime/C_n$. 
The PV appears to be suppressed exponentially as $n$ increases and  
one can easily choose the degree $n$ such that no PV would be observed within desired statistics.
Within our statistics,  no negative value of  $C_n^\prime/C_n$ was observed for $n>22$, 24 and 30 
at $\kappa=0.156$, 0.158 and 0.160 respectively.
However, these values of $n$ will depend on statistics: 
with high-statistics one may observe a small number of negative values even at large $n$. 
In order to remove this dependency, 
let us fix the rate of positivity violation.
In this study we are dealing with statistics of $O(10^3-10^4)$ trajectories.
So we set the PV level to $10^{-4}$.
From Fig.\ref{Sqb530nf1+1bw}, roughly speaking, 
we find that this level corresponds to $n\approx 23, 28, 34$
for $\kappa=0.156$, 0.158 and 0.160 respectively.
The lines indicated by "No Positivity Violation" 
in Figs.\ref{b530k156nf1+1w1x1bw}-\ref{b530k160nf1+1w2x2bw} correspond to these values of $n$.
In turn, as seen in Fig.\ref{b530L8nf1+1ACbw}, these values of $n$ 
correspond to an acceptance $\sim 95\%$ in Step 3 of our algorithm. 
This suggests that typically one needs an acceptance of 95\% or higher 
to maintain positivity of $C_n^\prime/C_n$ at the level of $10^{-4}$.
In terms of $C_n^\prime/C_n$, roughly speaking, 
95\% acceptance corresponds to $<|C_n^\prime/C_n-1|>\approx 0.1$.
Under these conditions, the contribution of configurations with negative
$C_n^\prime/C_n$ to the total ensemble, if present at all, will be 
well below the statistical fluctuations, and the way they are dealt with
is unimportant. We simply reject such configurations.

\section{Numerical results for $n_f=1$}
\begin{figure}[ht]
\vspace{-3cm}
  \begin{center}
    \epsfig{width=10cm,file=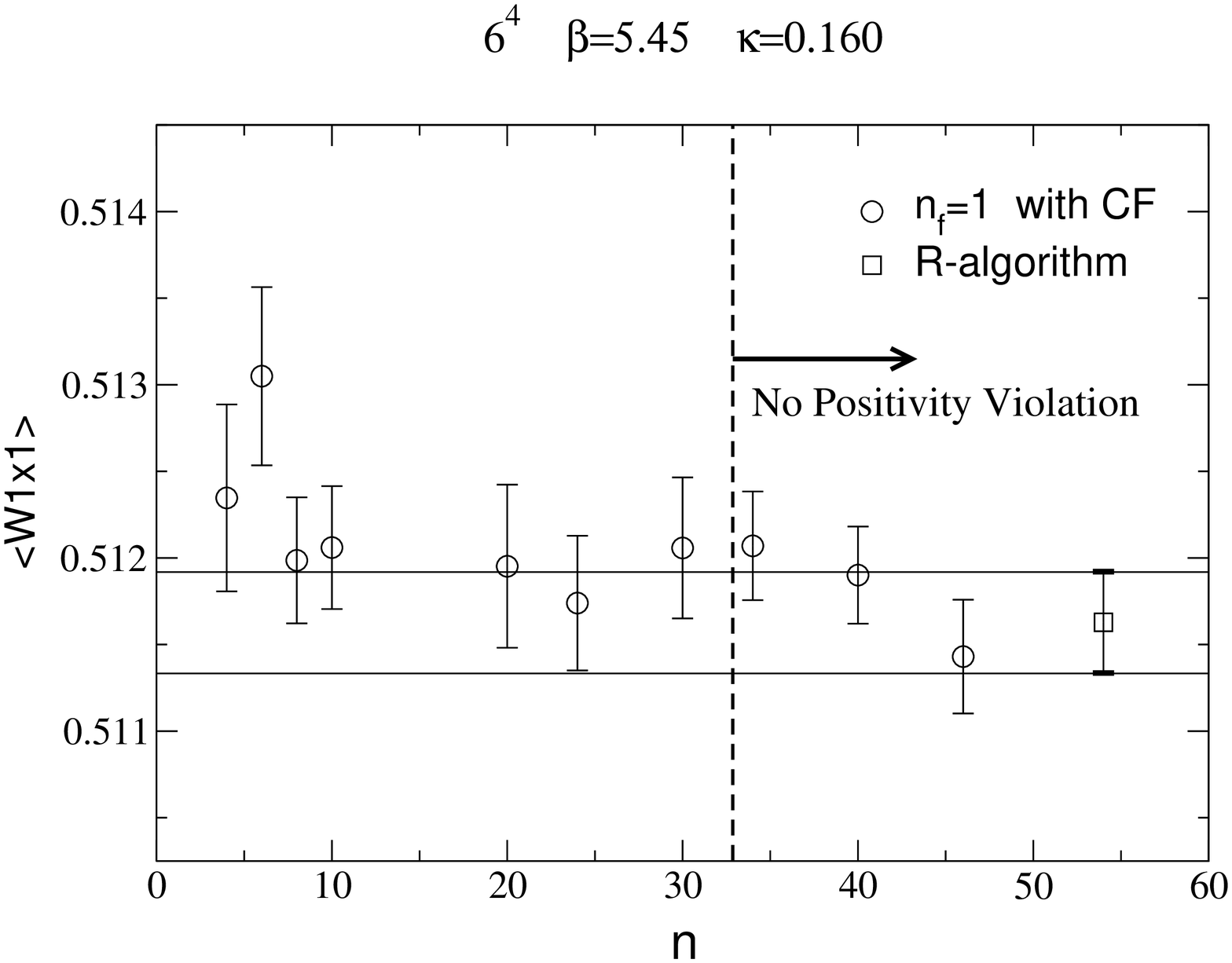}
    \epsfig{width=10cm,file=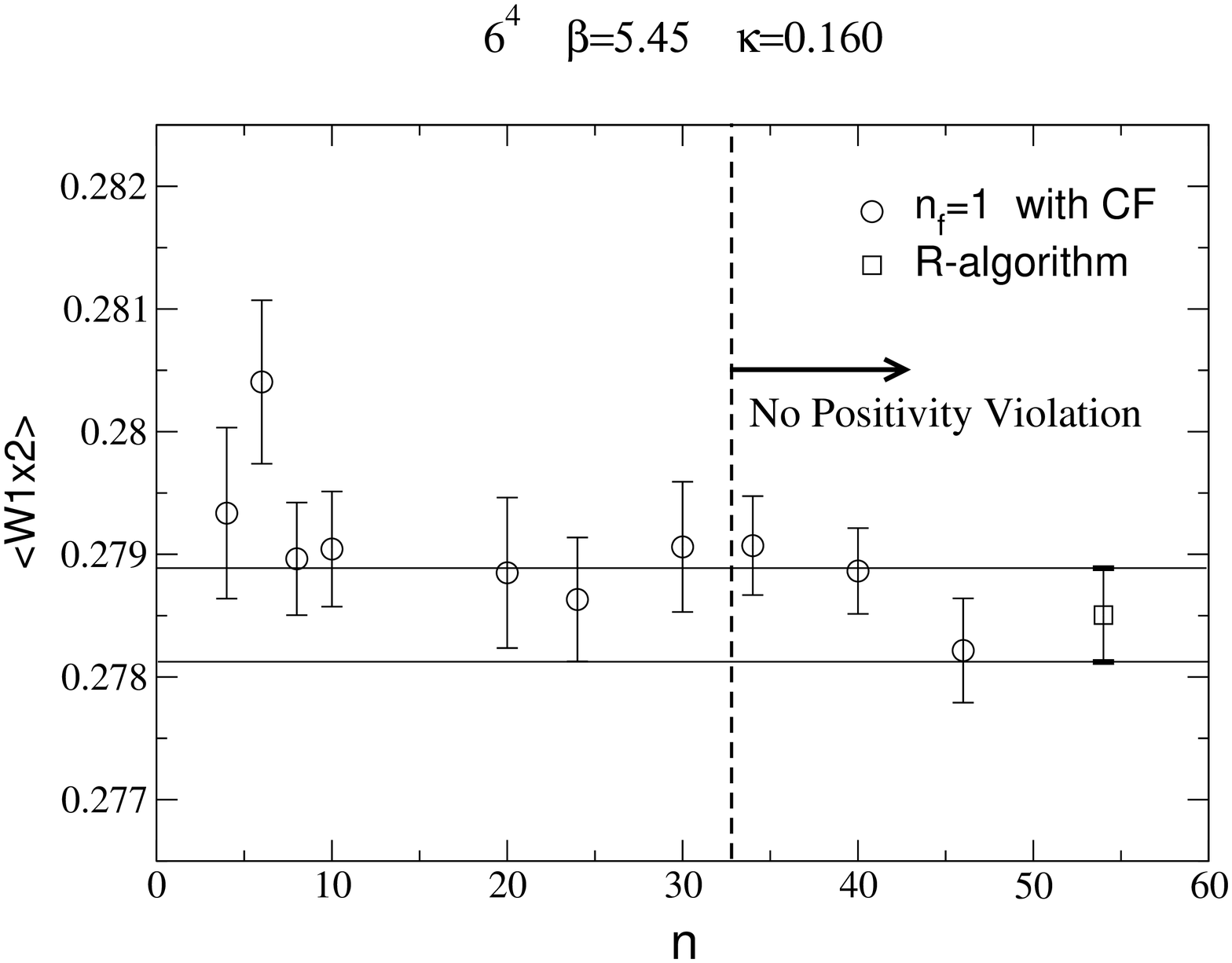}
    \epsfig{width=10cm,file=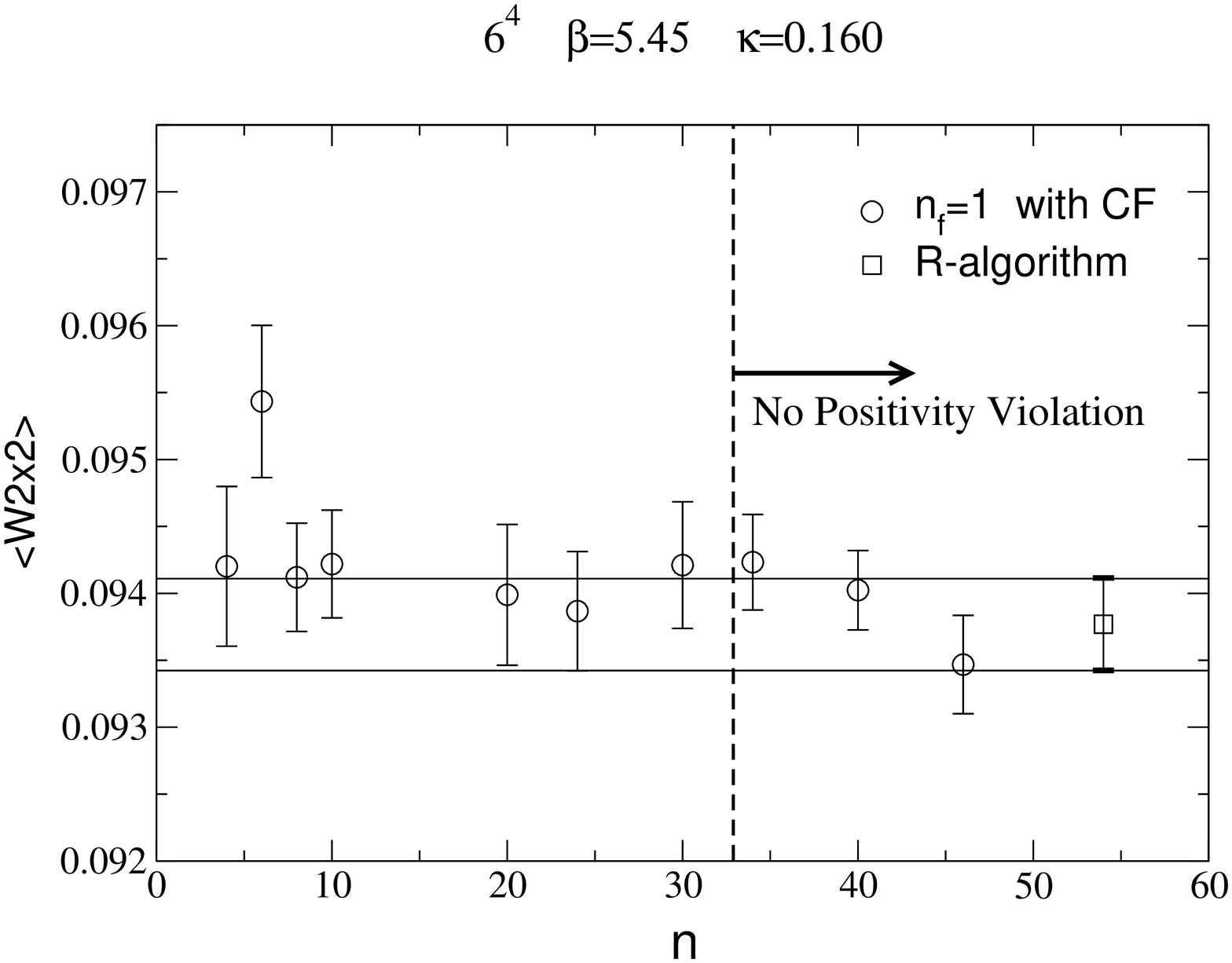}

  \end{center}
\vspace{-1cm}
\caption{$n_f=1$ results of Wilson loop on a $6^4$ lattice at $\beta=5.45$ and $\kappa=0.160$.
}
\label{b545L6k160w1x1bw}
\end{figure}

We take a $6^4$ lattice at $\beta=5.45$ and $\kappa=0.160$ 
with periodic boundary conditions in all directions. 
Simulations are performed using the algorithm of Sec.\ref{sec:nf1}.
Average values of Wilson loops are shown in Fig.\ref{b545L6k160w1x1bw}
together with results from the R-algorithm extrapolated to zero step-size.
The average values are taken over 10000-36000 trajectories. See Table \ref{paranf1} for details.
Our results are consistent with those from the R-algorithm except for small $n$.

\begin{figure}[ht]
  \begin{center}
    \epsfig{width=11cm,file=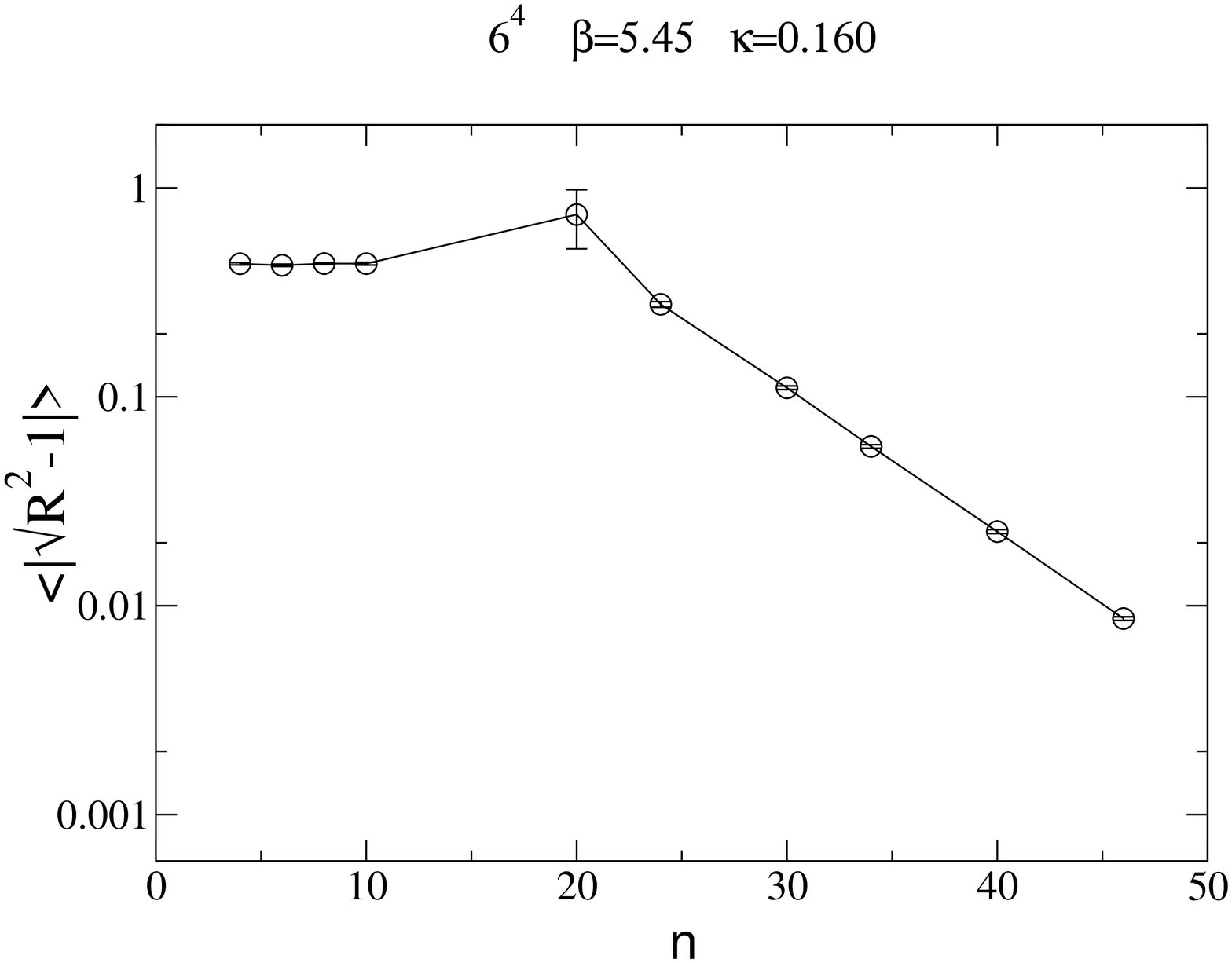}
  \end{center}
\vspace{-1cm}
\caption{$|C_n^\prime/C_n-1|$ for $n_f=1$ as a function of degree $n$.
$R$ in the figure stands for $C_n^\prime/C_n$.
}
\label{Sqb545L6k160}
\end{figure}

Fig.\ref{Sqb545L6k160} shows the average value of $|C_n^\prime/C_n-1|$.
Again, we see exponential convergence as $n$ increases, except for the plateau at small $n$. 

\begin{figure}[ht]
  \begin{center}
    \epsfig{width=11cm,file=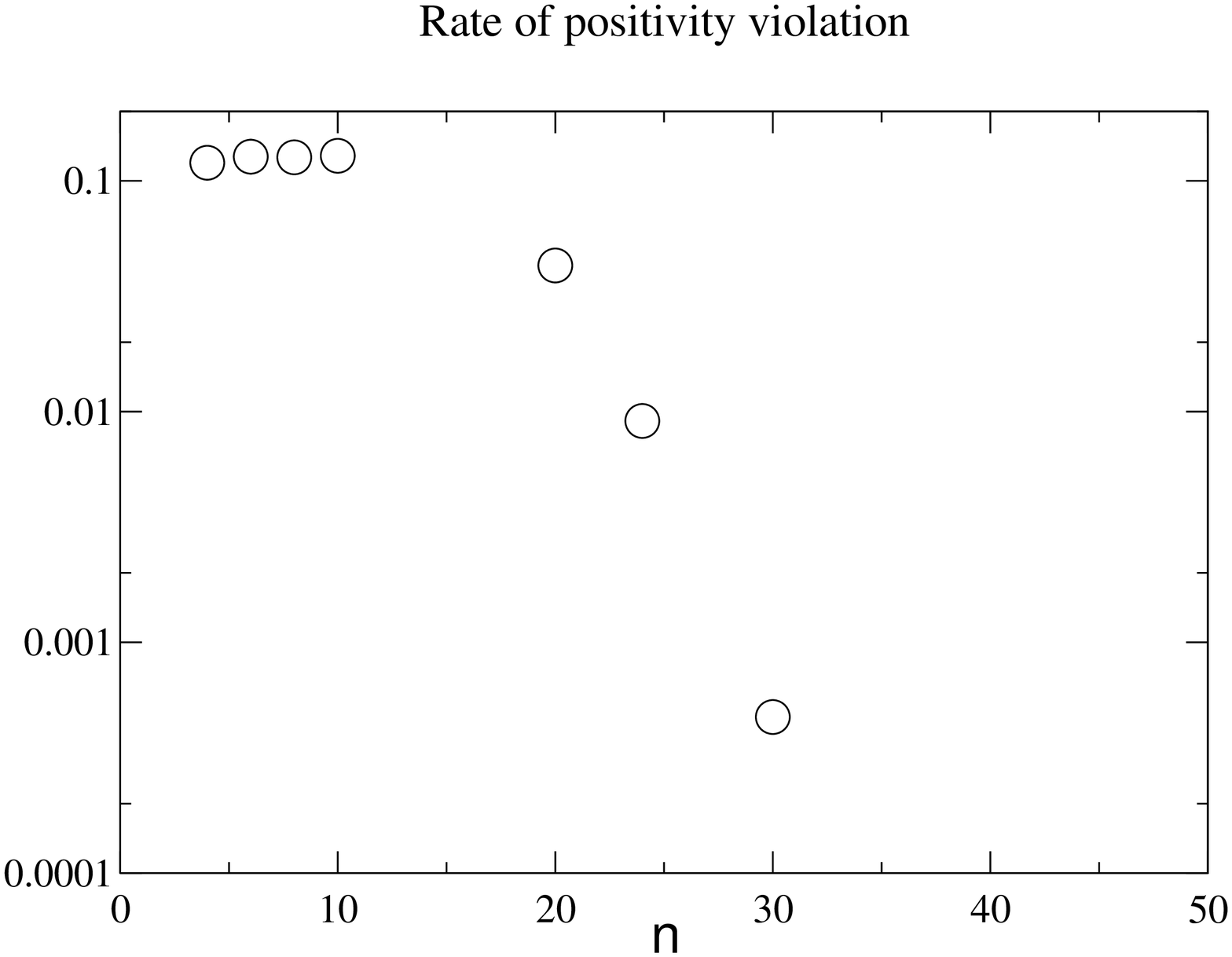}
  \end{center}
\vspace{-1cm}
\caption{Rate of positivity violation of $C_n^\prime/C_n$ for $n_f=1$ as a function of degree $n$.}
\label{b545L6k160PVbw}
\end{figure}

Fig.\ref{b545L6k160PVbw} shows the PV rate of $C_n^\prime/C_n$.
Positivity violation appears to be suppressed exponentially as $n$ increases.
Within our statistics, positivity was maintained for $n>30$. 
As discussed in Sec.\ref{sec:NRnf1+1}, adopting a PV level of $10^{-4}$, 
we find that this level corresponds to $n\approx 33$.

\begin{figure}[ht]
  \begin{center}
    \epsfig{width=11cm,file=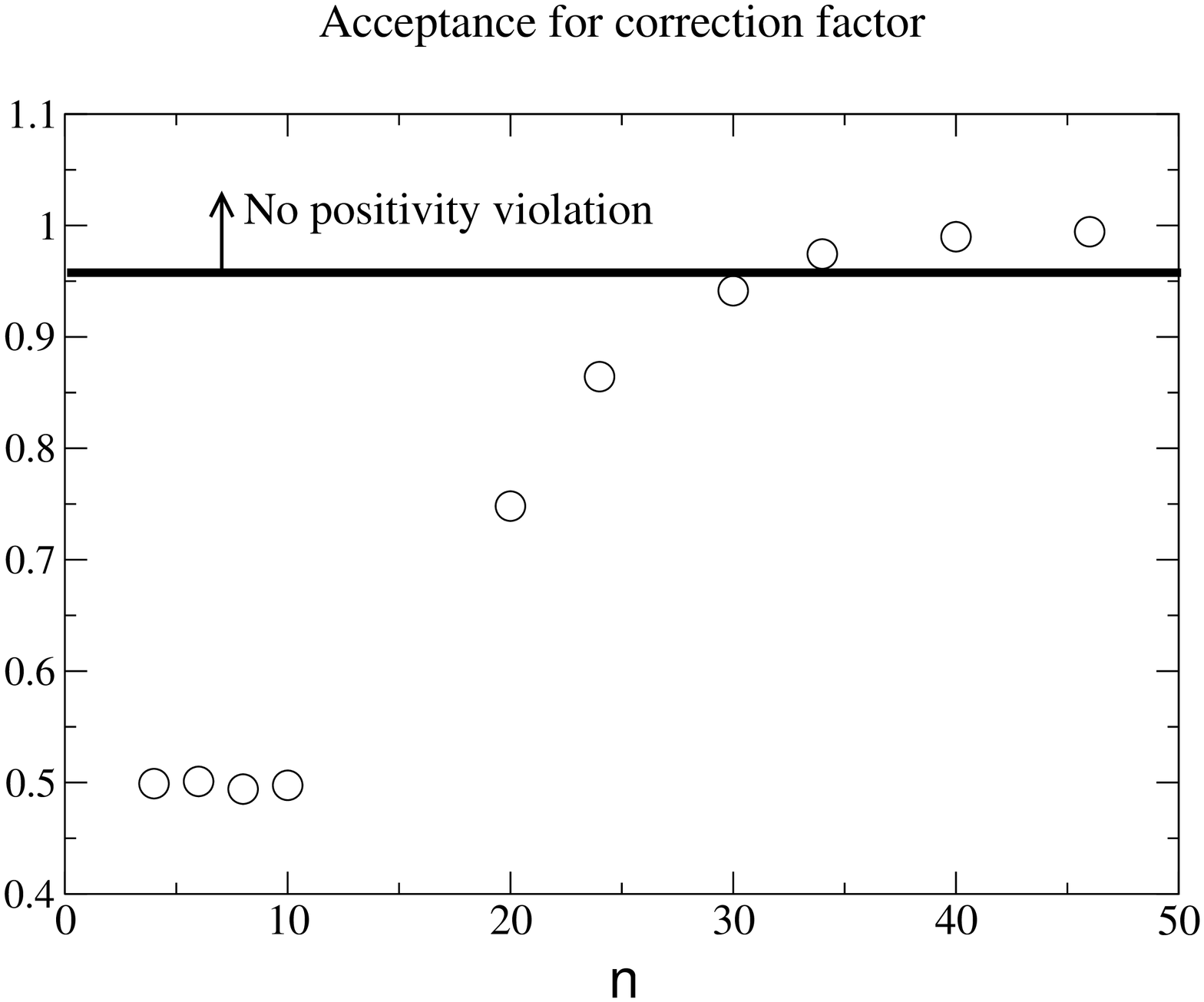}
  \end{center}
\vspace{-1cm}
\caption{Acceptance of Metropolis test with $p=\min(1,C_n^\prime/C_n)$ for $n_f=1$. }
\label{b545L6k160ACbw}
\end{figure}

Fig.\ref{b545L6k160ACbw} shows the average acceptance of the Metropolis test Step 3.
The PV level of $10^{-4}$ corresponds again to about 95\% acceptance,
and to $<|C_n^\prime/C_n-1|>\approx 0.07$.

\section{Numerical results for $n_f=2+1$}
We take a $6^4$ lattice at $\beta=5.30$ and $\kappa=0.156$
with periodic boundary conditions in all directions.
Simulations are performed using the algorithm of Sec.\ref{sec:nf2+1}.
Setting $\kappa\equiv \kappa_k =\kappa_l$, we can simulate 
$n_f=3$ QCD and compare results with those from $n_f=3$ R-algorithm.
The average values are taken over 9000-16000 trajectories. See Table \ref{paranf3} for details.
Our results are consistent with those from the R-algorithm extrapolated to 
zero step-size, except for small $n$.

\begin{figure}[ht]
  \begin{center}
    \epsfig{width=11cm,file=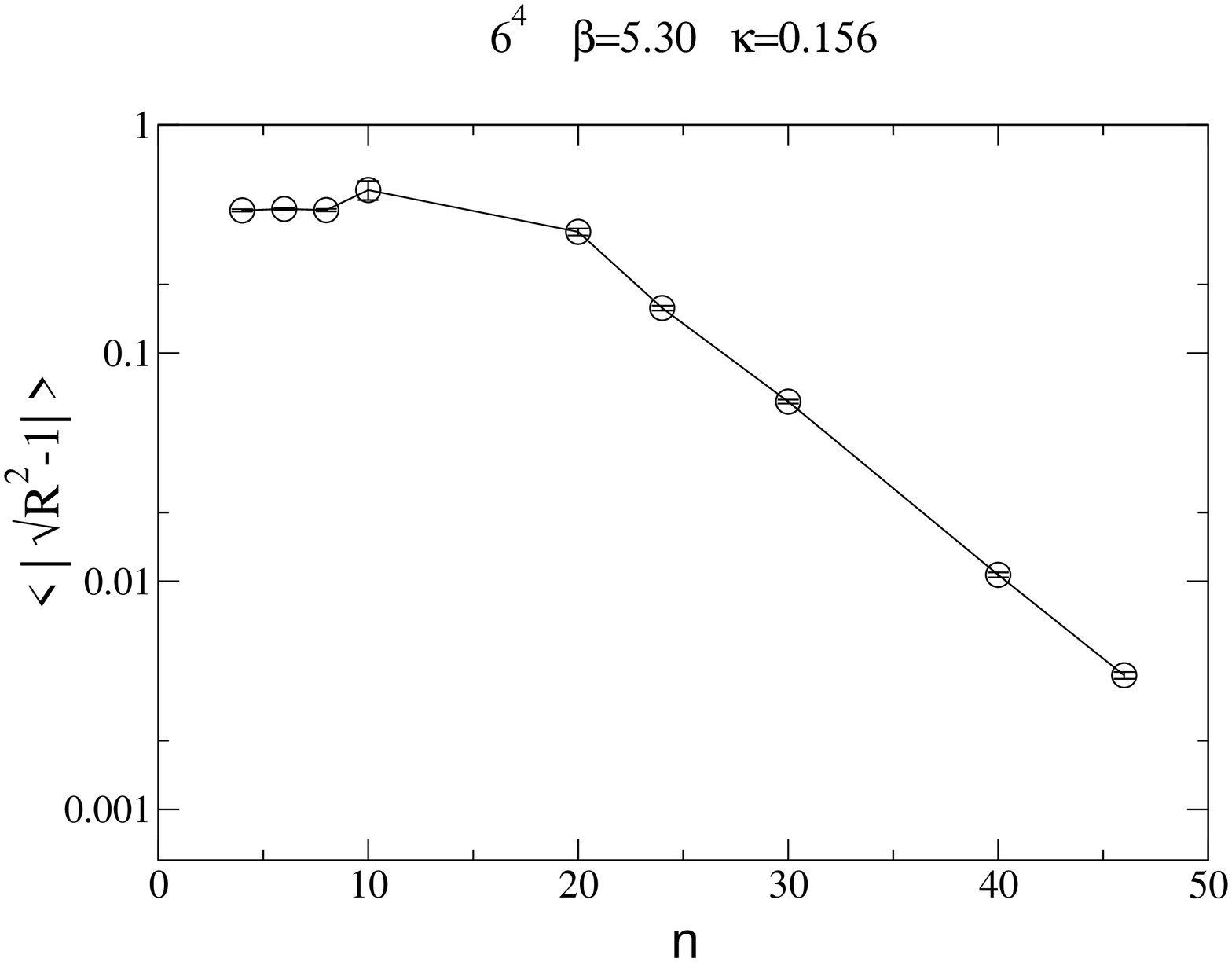}
  \end{center}
\vspace{-1cm}
\caption{$|C_n^\prime/C_n-1|$ for $n_f=2+1$ as a function of degree $n$.
$R$ in the figure stands for $C_n^\prime/C_n$.
}
\label{Sqb530nf3k156}
\end{figure}

Fig.\ref{Sqb530nf3k156} shows the average value of $|C_n^\prime/C_n-1|$.
Again we observe exponential convergence as $n$ increases, except for small $n$.                     

\begin{figure}[ht]
  \begin{center}
    \epsfig{width=11cm,file=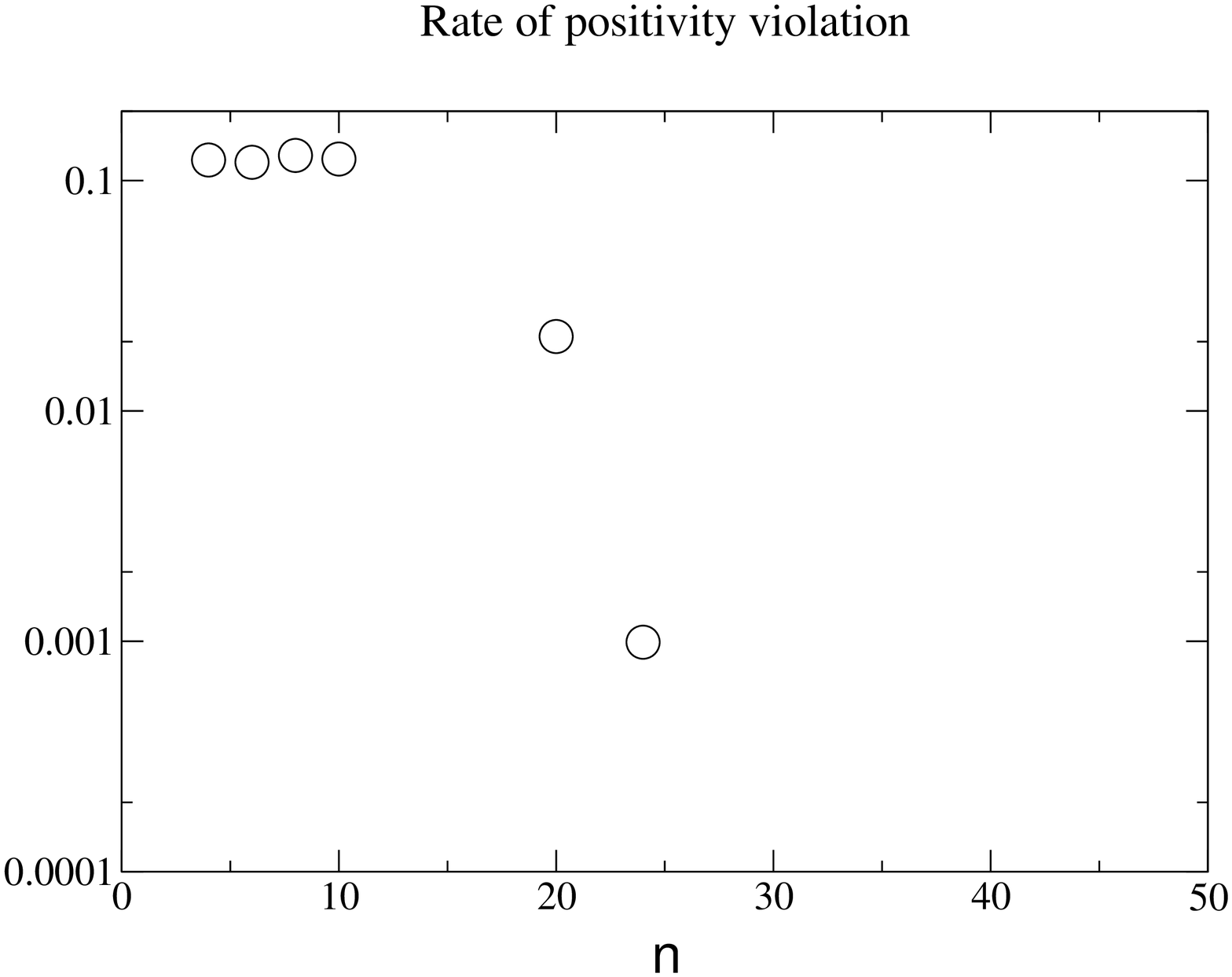}
  \end{center}
\vspace{-1cm}
\caption{Rate of positivity violation of $C_n^\prime/C_n$ for $n_f=2+1$ as a function of degree $n$.}
\label{b530nf3k156PV}
\end{figure}

Fig.\ref{b530nf3k156PV} shows the positivity-violation (PV) rate of $C_n^\prime/C_n$.
Within our statistics, positivity was maintained for $n>24$.
As discussed in Sec.\ref{sec:NRnf1+1}, adopting a PV level of $10^{-4}$,
we find that this level corresponds to $n\approx 27$.

\begin{figure}[ht]
  \begin{center}
    \epsfig{width=11cm,file=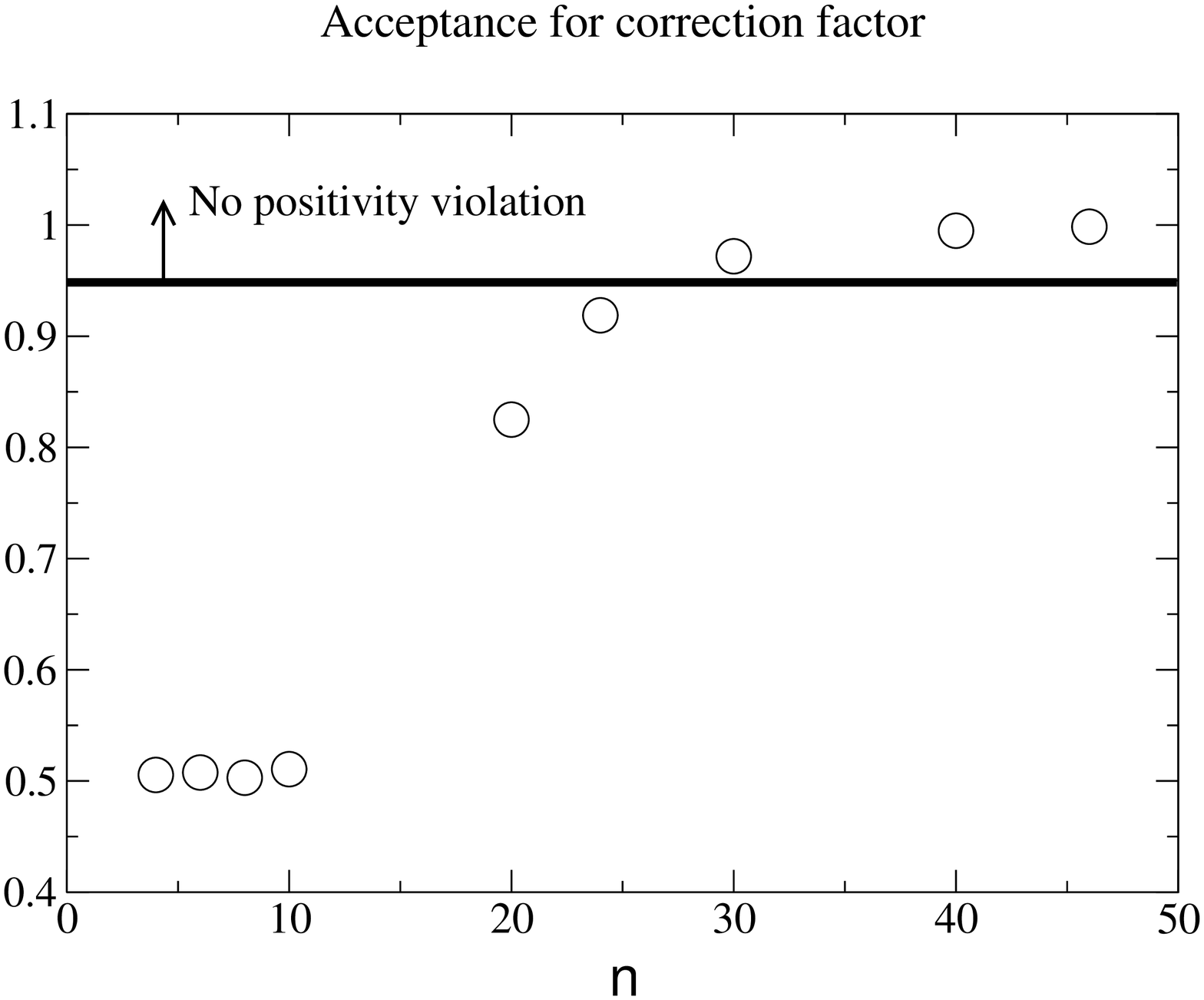}
  \end{center}
\vspace{-1cm}
\caption{Acceptance of Metropolis test with $p=\min(1,C_n^\prime/C_n)$ for $n_f=2+1$. }
\label{b530nf3k156CFac}
\end{figure}

Fig.\ref{b530nf3k156CFac} shows the average acceptance of the Metropolis test Step 3.
A PV level of $10^{-4}$ corresponds to about 95\% acceptance
and to $<|C_n^\prime/C_n-1|>\approx 0.1$.

\section{Conclusions}

We have discussed HMC algorithms for odd-flavor QCD simulations,
based on a polynomial approximation of the inverse Dirac operator $D^{-1}$.
The algorithms can be made exact with a correction factor 
which can be easily incorporated in an additional, stochastic Metropolis test.
We have tested the algorithms for $n_f=1,1+1$ and 2+1 flavors.
The results are consistent with those from the standard HMC and R-algorithm.

The estimator of the correction factor $C_n^\prime/C_n$ should be positive 
if it is to be used in a Metropolis test.
We observe that positivity violation is suppressed exponentially as $n$ increases.
Therefore, one can choose in advance a value of $n$ which will suffice to avoid
negative correction factors in the simulation for the desired statistics. 
When one fixes the positivity-violation level to $10^{-4}$, this corresponds to 
about $<|C_n^\prime/C_n-1|>\approx 0.07-0.1$.
Equivalently, this level corresponds to  
$\approx 95\%$ acceptance in the Metropolis test with $p=\min(1,C_n^\prime/C_n)$.

\section*{Acknowledgments}
The simulations were performed on the NEC SX-5 at INSAM, Hiroshima University and RCNP, Osaka University. 
About 1000 processor-hours were used.
This work was supported by the Ministry of Education, Science,
Sports and Culture, Grant-in-Aid, No.13740164.



\begin{table}
\caption{Step-size $\Delta \tau$, number of trajectories and acceptances
for $n_f=1+1$ QCD simulations on an $8^4$ lattice at $\beta=5.30$.
Acc.(HMC) stands for the acceptance with $p=\min(1,\exp(-\delta H))$ 
which corrects the integration step-size error, and
Acc.(CF) is the acceptance from the correction factor with
$p=\min(1,C_n^\prime/C_n)$, which corrects the error of the polynomial approximation.
}
\label{paranf1+1}
\begin{tabular}{cccccc} \hline

$\kappa$ & $n$ & $\Delta \tau$ & trajectories &   Acc.(HMC)\%  & Acc.(CF)\%  \\ \hline
0.156    &  4  & 0.05       &  3000        &    78          &    52   \\
         &  6  & 0.05       &  3000        &    80          &    51   \\
         &  8  & 0.05       &  3000        &    80          &    51   \\
         & 10  & 0.05       &  4500        &    80          &    49   \\
         & 20  & 0.05       &  3000        &    81          &    88   \\
         & 22  & 0.05       &  4000        &    81          &    93   \\
         & 24  & 0.05       &  3000        &    80          &    96   \\
         & 30  & 0.05       &  4000        &    81          &    98.9   \\
         & 40  & 0.05       &  4200        &    82          &    100   \\
         & 46  & 0.05       &  3600        &    80          &    100   \\ \\

0.158    &  4  & 0.05       &  3000        &    78          &    49   \\
         &  6  & 0.05       &  4500        &    77          &    50   \\
         &  8  & 0.05       &  3000        &    80          &    50   \\
         & 10  & 0.05       &  4500        &    80          &    49   \\
         & 20  & 0.05       &  5000        &    81          &    77   \\
         & 24  & 0.05       &  4000        &    80          &    90   \\
         & 30  & 0.05       &  4000        &    80          &    98   \\
         & 40  & 0.05       &  4900        &    80          &    99.7   \\
         & 46  & 0.05       &  4500        &    80          &    99.9   \\ \\

0.160    &  4  & 0.05       &  5000        &    78          &    51   \\
         &  6  & 0.05       &  4000        &    79          &    50   \\
         &  8  & 0.05       &  7500        &    80          &    50   \\
         & 10  & 0.05       &  4500        &    80          &    49   \\
         & 20  & 0.05       &  6000        &    79          &    61   \\
         & 24  & 0.05       &  3600        &    80          &    77   \\
         & 30  & 0.05       &  3200        &    81          &    92   \\
         & 40  & 0.05       &  2800        &    81          &    98.8   \\
         & 46  & 0.05       &  3000        &    81          &    99.7   \\

\hline
\end{tabular}
\end{table}

\begin{table}
\caption{$n_f=1+1$ simulation results for $1\times 1$, $1\times 2$ and $2\times 2$ Wilson loops
on an $8^4$ lattice at $\beta=5.30$.}

\begin{tabular}{ccccc} \hline
$\kappa$  &  $n$ & $1\times 1$  &  $1\times 2$ & $2\times 2$ \\  \hline
0.156   &     4  & 0.48846(37)  &  0.25172(48) & 0.07455(37)  \\
        &     6  & 0.48637(28)  &  0.24927(34) & 0.07276(22)  \\
        &     8  & 0.48502(44)  &  0.24780(51) & 0.07176(39)  \\
        &    10  & 0.48566(48)  &  0.24845(60) & 0.07235(45)  \\
        &    20  & 0.48548(39)  &  0.24824(45) & 0.07212(32)  \\
        &    22  & 0.48568(16)  &  0.24853(20) & 0.07234(16)  \\
        &    24  & 0.48589(34)  &  0.24876(42) & 0.07246(32)  \\
        &    30  & 0.48566(25)  &  0.24845(31) & 0.07224(23)  \\
        &    40  & 0.48587(31)  &  0.24877(40) & 0.07245(31)  \\
        &    46  & 0.48613(28)  &  0.24906(36) & 0.07271(26)  \\
        &  HMC   & 0.48594(33)  &  0.24883(41) & 0.07256(32)  \\  \\  
0.158   &     4  & 0.49274(48)  &  0.25698(58) & 0.07843(46)  \\
        &     6  & 0.49129(44)  &  0.25531(57) & 0.07730(48)  \\
        &     8  & 0.49032(48)  &  0.25403(55) & 0.07628(37)  \\
        &    10  & 0.49055(46)  &  0.25431(57) & 0.07646(40)  \\
        &    20  & 0.49020(24)  &  0.25391(28) & 0.07615(19)  \\
        &    24  & 0.49055(29)  &  0.25439(35) & 0.07661(26)  \\
        &    30  & 0.49003(39)  &  0.25373(47) & 0.07603(36)  \\
        &    40  & 0.49027(42)  &  0.25401(52) & 0.07634(40)  \\
        &    46  & 0.49009(21)  &  0.25378(25) & 0.07608(20)  \\
        &  HMC   & 0.49026(20)  &  0.25397(25) & 0.07624(19)  \\  \\
0.160   &    4  & 0.49814(42)  &  0.26347(52) & 0.08326(42)  \\
        &    6  & 0.49604(48)  &  0.26110(61) & 0.08180(49)  \\
        &    8  & 0.49546(55)  &  0.26030(69) & 0.08102(54)  \\
        &    10  & 0.49515(34)  &  0.25982(45) & 0.08054(40)  \\
        &    20  & 0.49457(34)  &  0.25923(42) & 0.08016(34)  \\
        &    24  & 0.49569(29)  &  0.26055(35) & 0.08120(30)  \\
        &    30  & 0.49555(33)  &  0.26038(49) & 0.08102(34)  \\
        &    40  & 0.49564(41)  &  0.26051(51) & 0.08106(40)  \\
        &    46  & 0.49529(43)  &  0.26014(55) & 0.08089(42)  \\
        &  HMC   & 0.49573(48)  &  0.26068(62) & 0.08131(49)  \\

\hline
\end{tabular}
\end{table}

\vspace{1cm}

\begin{table}
\caption{Same as in table \ref{paranf1+1}, but for $n_f=1$ QCD simulations
on a $6^4$ lattice at $\beta=5.45$.}
\label{paranf1}
\begin{tabular}{cccccc} \hline

$\kappa$ & $n$ & $\Delta \tau$ & trajectories &   Acc.(HMC)\%  & Acc.(CF)\%
\\ \hline
0.160    &  2  & 0.04       &  5000         &    93          &    49   \\
         &  4  & 0.04       &  22000        &    93          &    50   \\
         &  6  & 0.04       &  36000        &    92          &    50   \\
         &  8  & 0.04       &  20000        &    93          &    49   \\
         & 10  & 0.05       &  15000        &    89          &    50   \\
         & 20  & 0.04       &  14000        &    93          &    75   \\
         & 24  & 0.04       &  10500        &    93          &    86   \\
         & 30  & 0.04       &  14300        &    93          &    94   \\
         & 34  & 0.05       &  16500        &    89          &    97   \\
         & 40  & 0.05       &  15000        &    89          &    99   \\
         & 46  & 0.05       &  22000        &    89          &    99.4   \\
\\
\hline
\end{tabular}
\end{table}

\vspace{1cm}

\begin{table}
\caption{$n_f=1$ simulation results for $1\times 1$, $1\times 2$ and
$2\times 2$ Wilson loops
on a $6^4$ lattice at $\beta=5.45$.
Results from the R-algorithm were extrapolated to step-size $\Delta \tau=0$,
based on
simulation results at $\Delta \tau=0.0125$, 0.025, 0.0333333 and 0.04.
}

\begin{tabular}{ccccc} \hline
$\kappa$  &  $n$ & $1\times 1$  &  $1\times 2$ & $2\times 2$ \\  \hline
0.160     &   2 & 0.49637(60)  & 0.25899(73)  & 0.07794(59) \\
          &   4 & 0.51235(54)  & 0.27934(70)  & 0.09420(60) \\
          &   6 & 0.51305(51)  & 0.28041(67)  & 0.09543(57) \\
          &   8 & 0.51199(36)  & 0.27896(46)  & 0.09412(40) \\
          &   10 & 0.51206(36)  & 0.27904(47)  & 0.09422(40) \\
          &   20 & 0.51195(47)  & 0.27885(61)  & 0.09399(53) \\
          &   24 & 0.51174(39)  & 0.27863(51)  & 0.09387(45) \\
          &   30 & 0.51206(41)  & 0.27906(53)  & 0.09421(47) \\
          &   34 & 0.51207(31)  & 0.27907(40)  & 0.09423(36) \\
          &   40 & 0.51190(28)  & 0.27886(35)  & 0.09402(30) \\
          &   46 & 0.51143(33)  & 0.27822(43)  & 0.09347(37) \\
   & R-algorithm & 0.51163(29)  & 0.27850(38)  & 0.09377(35) \\
\hline
\end{tabular}
\end{table}

\begin{table}
\caption{Same as in table \ref{paranf1+1} but for $n_f=2+1$ QCD simulations
on a $6^4$ lattice at $\beta=5.30$.}
\label{paranf3}
\begin{tabular}{cccccc} \hline

$\kappa$ & $n$ & $\Delta \tau$ & trajectories &   Acc.(HMC)\%  & Acc.(CF)\%
\\ \hline
0.156    &  2  & 1/18       &  9000        &    80          &    50   \\
         &  4  & 1/18       &  12100        &    80          &    51   \\
         &  6  & 1/18       &  15700        &    81          &    51   \\
         &  8  & 1/18       &  12000        &    80          &    50   \\
         & 10  & 1/18       &  9000        &    79          &    51   \\
         & 20  & 1/18       &  12500        &    81          &    82   \\
         & 24  & 1/18       &  10000        &    81          &    92   \\
         & 30  & 1/18       &  12500        &    81          &    97   \\
         & 40  & 1/18       &  14000        &    81          &    99.5   \\
         & 46  & 1/18       &  16000        &    80          &    99.9   \\
\\
\hline
\end{tabular}
\end{table}

\vspace{1cm}

\begin{table}
\caption{$n_f=2+1$ simulation results for $1\times 1$, $1\times 2$ and
$2\times 2$ Wilson loops
on a $6^4$ lattice at $\beta=5.30$.
Results from the R-algorithm were extrapolated to step-size $\Delta \tau=0$,
based on
simulation results at $\Delta \tau=0.0125$, 0.0333333, 0.04 and 0.05.
}

\begin{tabular}{cclll} \hline
$\kappa$  &  $n$ & $1\times 1$  &  $1\times 2$ & $2\times 2$ \\  \hline
0.156     &   2 & 0.50032(46)  & 0.26559(58)  & 0.08444(48) \\
          &   4 & 0.51785(77)  & 0.2882(10)   & 0.10343(95) \\
          &   6 & 0.5206(11)   & 0.2919(15)   & 0.1071(14) \\
          &   8 & 0.52072(92)  & 0.2922(12)   & 0.1075(11) \\
          &   10 & 0.52005(69)  & 0.29124(92)  & 0.10648(85) \\
          &   20 & 0.52161(71)  & 0.29338(97)  & 0.10859(94) \\
          &   24 & 0.51993(77)  & 0.2911(10)   & 0.10639(99)\\
          &   30 & 0.52131(98)  & 0.2930(13)   & 0.1082(12) \\
          &   40 & 0.52009(75)  & 0.2913(10)   & 0.10657(98) \\
          &   46 & 0.51950(79)  & 0.2905(11)   & 0.1059(10) \\
   & R-algorithm & 0.5204(10)   & 0.2917(14)   & 0.1069(14) \\
\hline
\end{tabular}
\end{table}


\newpage

\begin{figure}[ht]
\vspace{-3cm}
  \begin{center}
    \epsfig{width=10cm,file=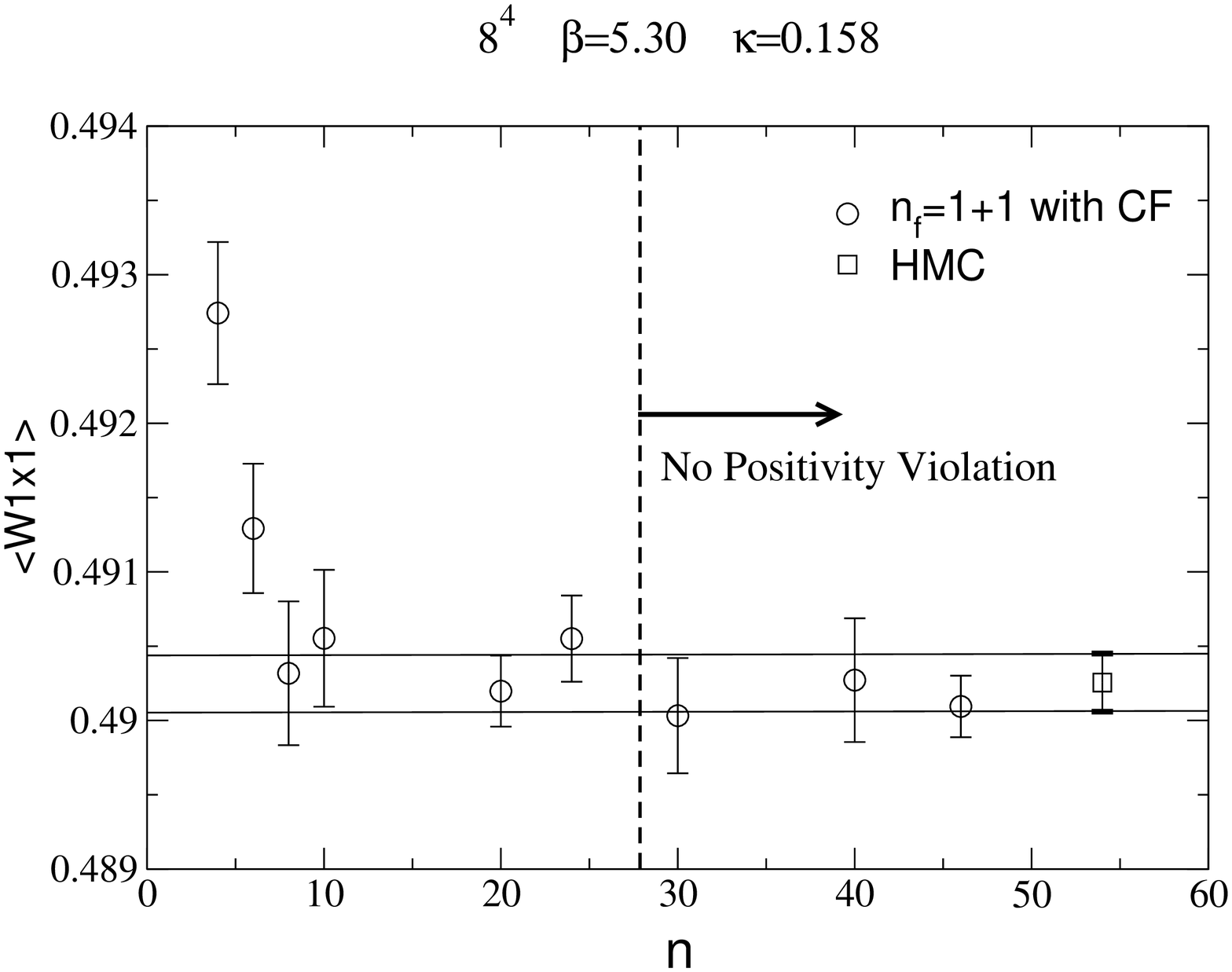}
    \epsfig{width=10cm,file=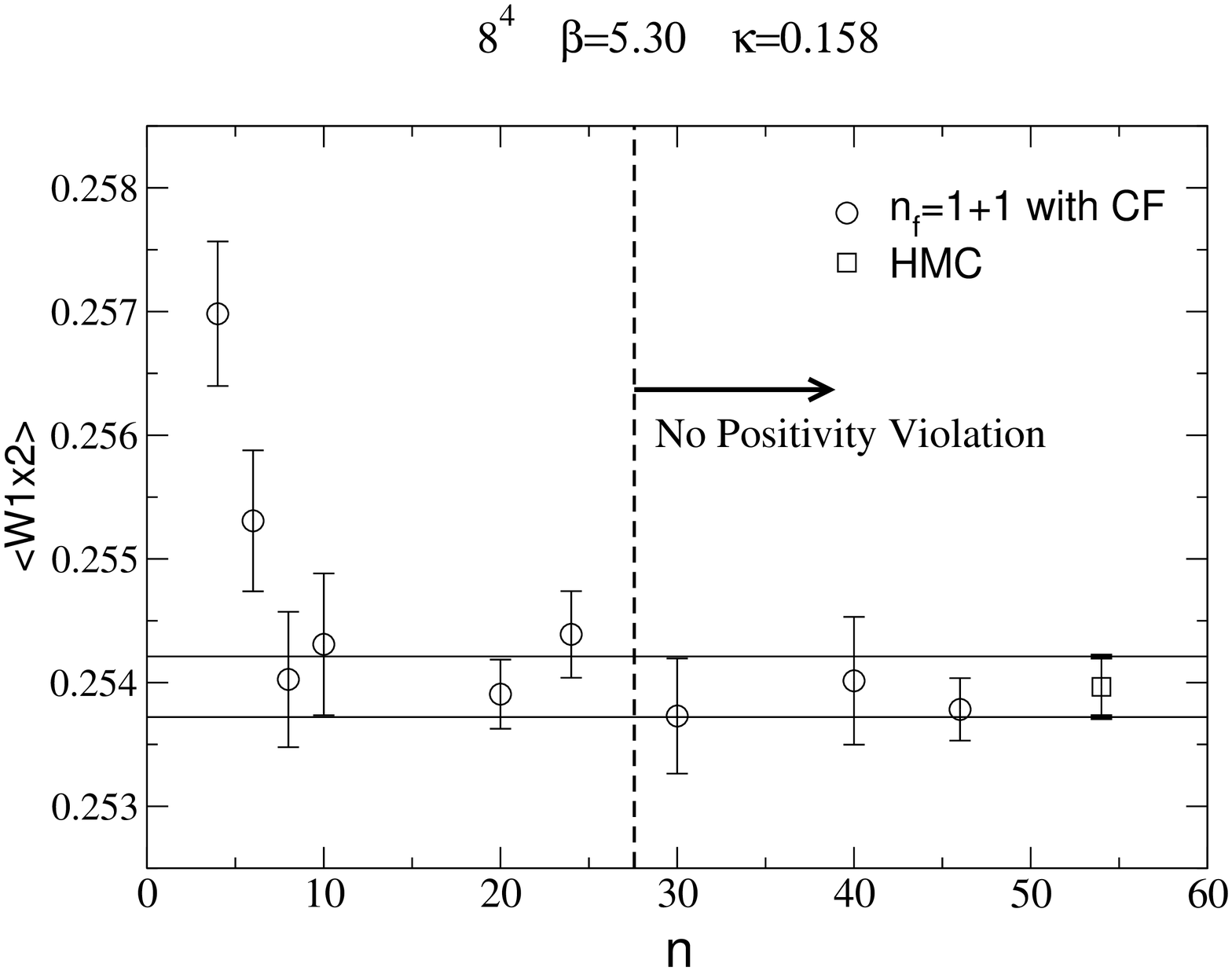}
    \epsfig{width=10cm,file=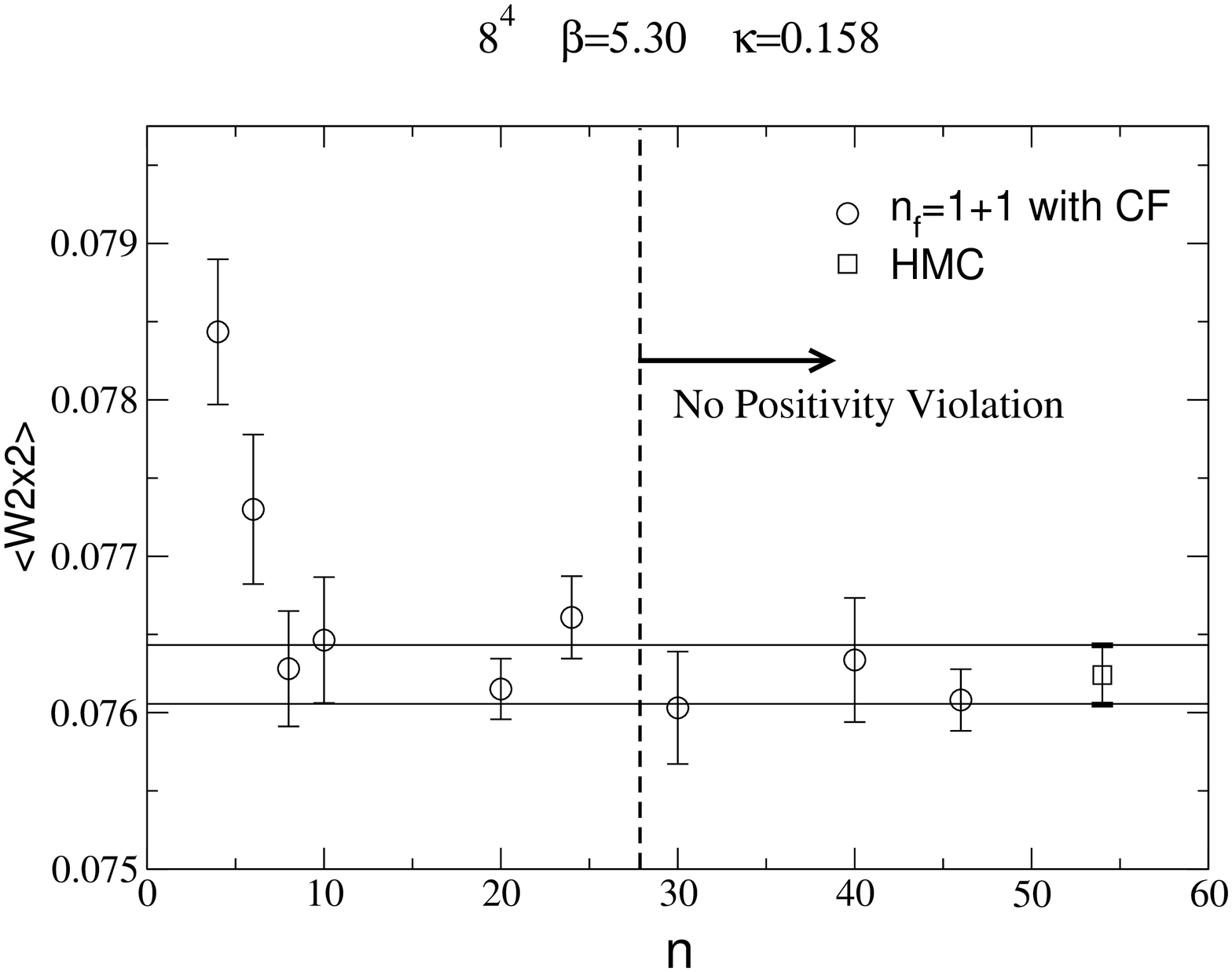}
  \end{center}
\vspace{-1cm}
\caption{Same as in fig.\ref{b530k156nf1+1w1x1bw} but for $\kappa=0.158$.}
\end{figure}

\newpage


\begin{figure}[ht]
\vspace{-3cm}
  \begin{center}
    \epsfig{width=10cm,file=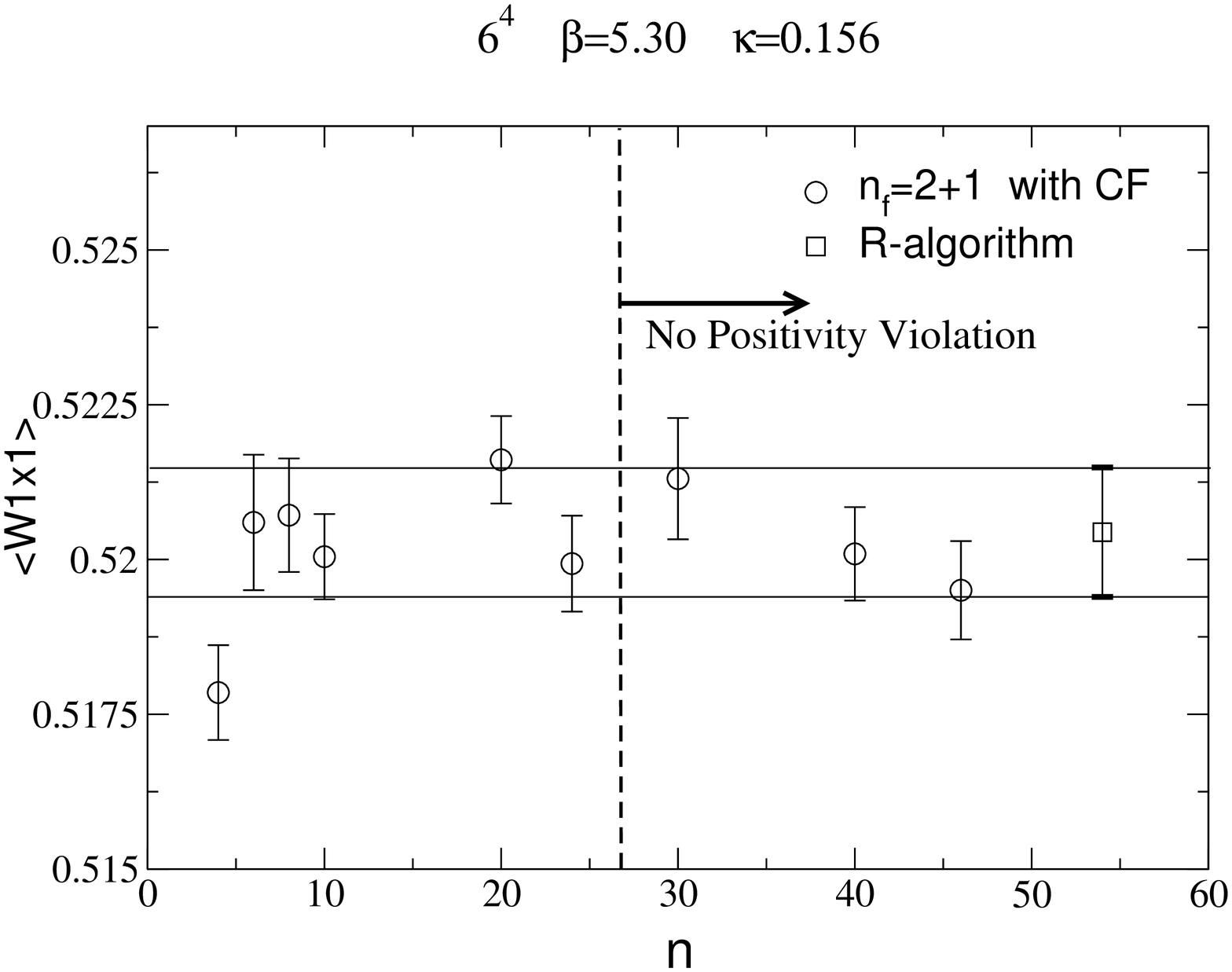}
    \epsfig{width=10cm,file=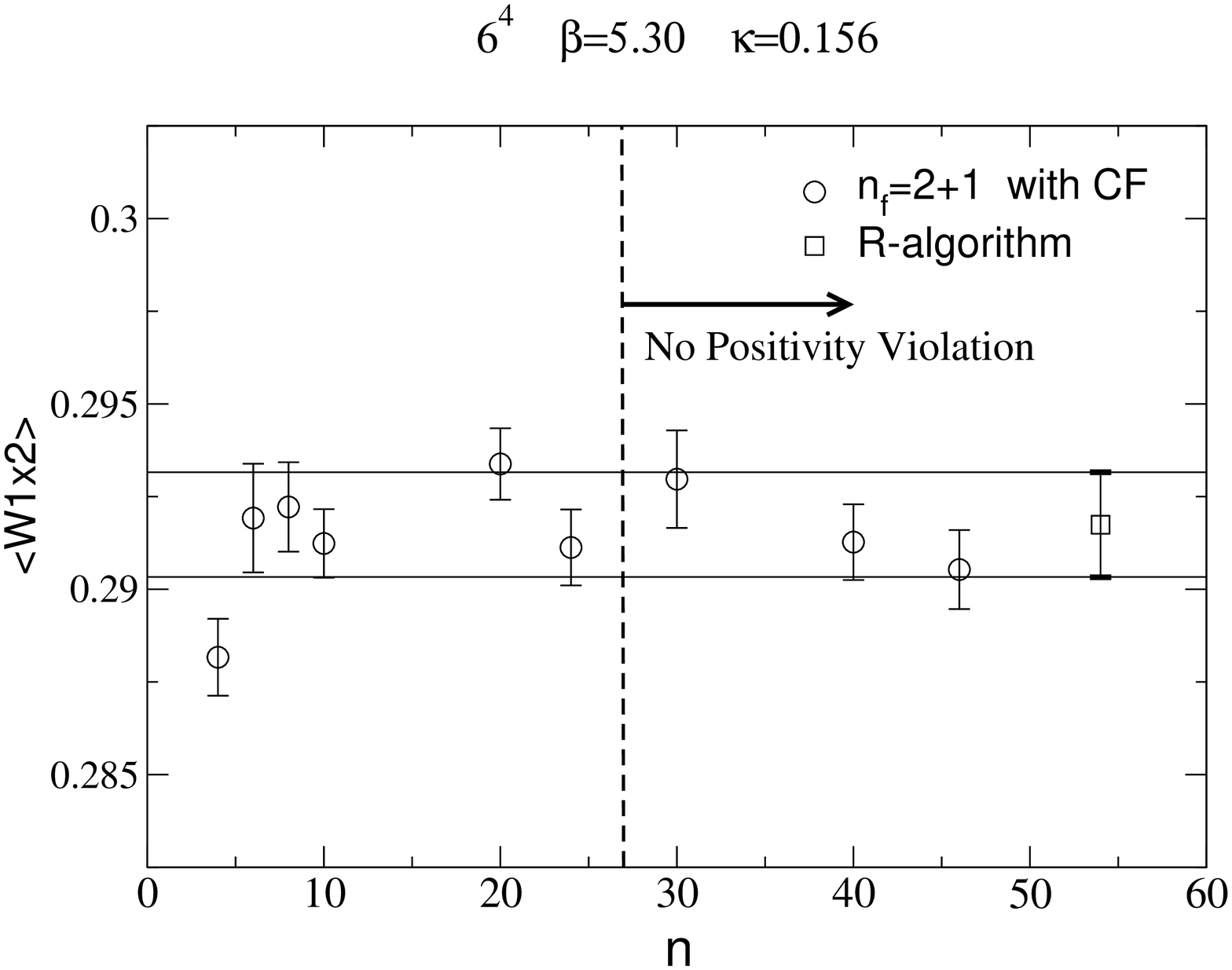}
    \epsfig{width=10cm,file=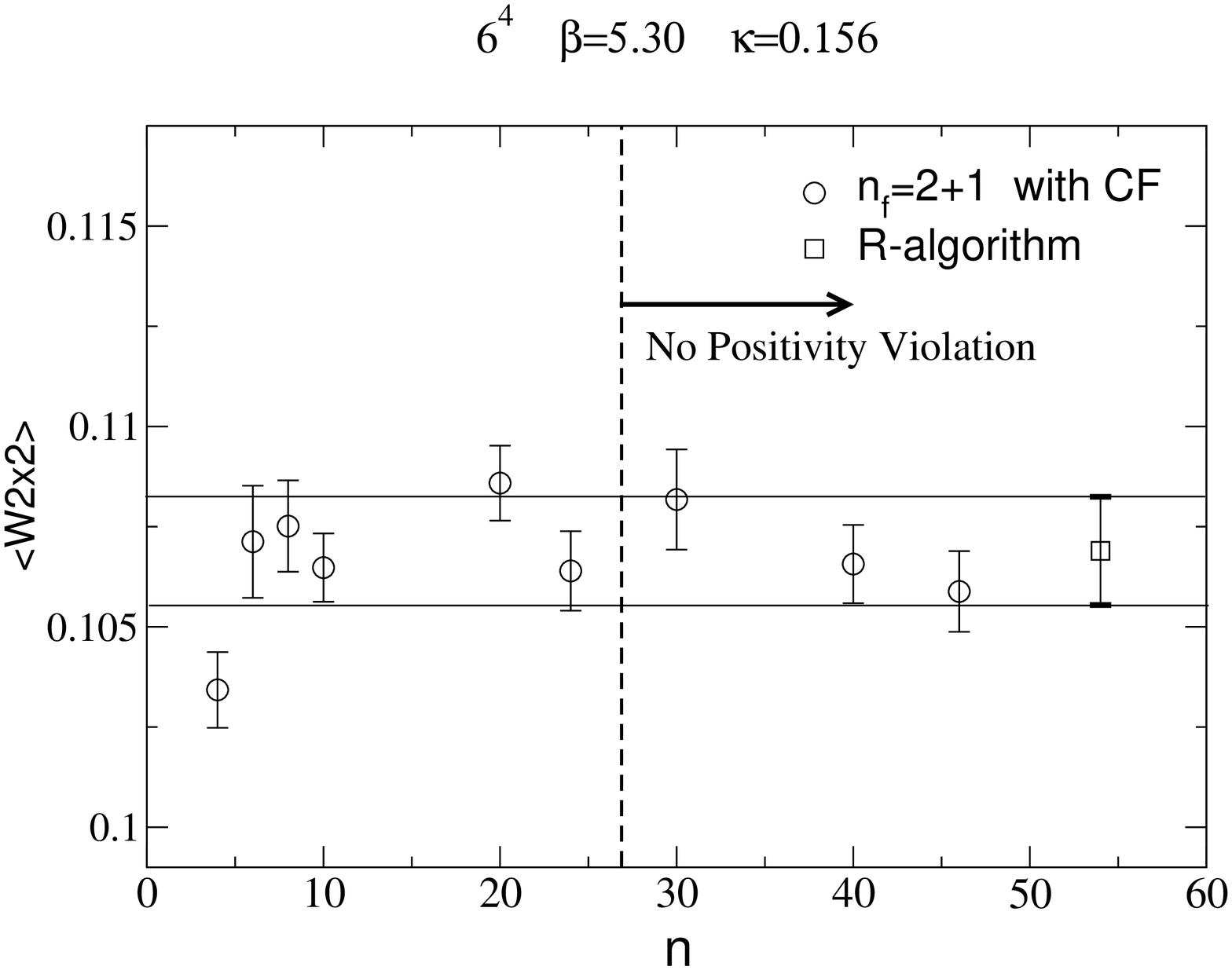}
  \end{center}
\vspace{-1cm}
\caption{$n_f=2+1$ results of Wilson loop on a $6^4$ lattice at $\beta=5.30$ and $\kappa=0.156$.}
\label{b530nf3k156w1x1bw}
\end{figure}

\end{document}